\documentclass[12pt]{article}
\usepackage{graphicx}
\usepackage{bm}
\begin{document}
\newcommand{\cd}{\makebox[0.08cm]{$\cdot$}}

\centerline{\bf $J/\psi$ production in an equilibrating partonic system}
\vskip 14pt
\centerline{Xiao-Ming Xu}
\vskip 14pt
\centerline{Institute for Nuclear Theory, University of Washington, Box
351550,}
\centerline{Seattle, WA 98195}
\centerline{and}
\centerline{Nuclear Physics Division, Shanghai Institute of Nuclear Research}
\centerline{Chinese Academy of Sciences, P.O.Box 800204, Shanghai 201800,
            China}

\begin{abstract}
\baselineskip=14pt
Any color singlet or octet $c\bar c$ pair is created at
short distances and then expands to a full size of $J/\psi$. Such a
dynamical evolution process is included here in calculations for the $J/\psi$
number distribution as a function of transverse momentum and rapidity in
central Au-Au collisions at both RHIC and LHC energies.
The $c\bar c$ pairs are produced in the initial collision and in the partonic
system during the prethermal
and thermal stages through the partonic channels $ab \to c\bar {c}[
^{2S+1}L_J]$ and $ab \to c\bar {c}[ ^{2S+1}L_J]{\rm x}$,
and then they dissociate in the latter two stages.
Dissociation of $c\bar c$ in the
medium occurs via two reactions: (a) color singlet $c\bar c$ plus a gluon turns
to color octet $c\bar c$ , (b) color octet $c\bar c$ plus a gluon persists as
color octet. There are modest yields of
$c\bar c$ in the prethermal stage at RHIC energy and through the reactions
$ab \to c\bar {c}[ ^{2S+1}L_J]$ at LHC energy for partons
with large average momentum in the prethermal stage at both collider
energies and in the thermal stage at LHC energy. Production from the
partonic system competes
with the suppression of the initial yield in the deconfined medium.
Consequently, a
bulge within $-1.5<y<1.5$ has been found for the $J/\psi$ number
distribution and the ratio of $J/\psi$ number distributions for Au-Au
collisions to nucleon-nucleon collisions. This
bulge is caused by the partonic system and is thus an indicator of a
deconfined
partonic medium. Based on this result we suggest the rapidity region worth
measuring in future experiments at RHIC and LHC to be $-3<y<3$.
\end{abstract}
\leftline{PACS codes: 24.85.+p, 12.38.Mh, 25.75.Dw, 25.75.Gz}
\leftline{Keywords: Ultrarelativistic nucleus-nucleus collisions, Equilibrating
partonic}
\leftline{~~~~~~~~~~~~~~~system, $J/\psi$ number distribution, Survival
probability}

\newpage
\leftline{\bf 1. Introduction}
\vspace{0.5cm}
A hot deconfined medium favors the dissociation of $J/\psi$ since
enough hard gluons can overcome the large energy gap between the $J/\psi$ and a
continuum state of $c\bar c$ [1]. Models based
on perturbative QCD have shown that a dense partonic system can be produced in
central Au-Au collisions at
RHIC and LHC energies [2-6] and then evolve toward thermal equilibrium and
likely chemical equilibrium [7-11]. Such parton plasmas will be searched for
soon in experiments at Brookhaven National Laboratory Relativistic Heavy Ion
Collider (RHIC). The $J/\psi$ suppression has been taken as
a thermometer to identify the evolution history of a parton plasma by
showing transverse momentum dependence of the survival probability in the
central rapidity region [12].

Charmonium melting inside a hot medium, which leads to $J/\psi$
suppression, was proposed by Matsui and Satz to probe the existence of the
quark-gluon plasma [13]. Before the complete formation of charmonium is
achieved, a pre-resonant $c\bar c$ is expanding from a collision point.
Dominance of the
color octet plus a collinear gluon configuration in the pre-resonance
state [14] may account for the same suppression of $\psi'$ and $J/\psi$
production in proton-nucleus collisions [15]. The growth of the color octet
configuration and its interaction with nucleons along its trajectory in a
nucleus are essential ingredients
in explaining measured $J/\psi$ production cross sections. In
addition, the importance of color octet configurations has been verified
in $p\bar {p}$ collisions at center-of-mass energy $\sqrt {s}=1.8$ TeV with
the CDF detector at Fermilab [16]. Theoretically, the color-octet production
at short distances and its evolution into physical resonances has been well
formulated in nonrelativistic QCD [17]. At the collision energies of RHIC and
LHC we can reasonably
expect considerable contributions from the color octet mechanism.

The evolution of ultrarelativistic nucleus-nucleus collisions, e.g. central
Au-Au collisions at both RHIC and LHC energies, has been divided
into three stages in Refs. [9,18,19]: (a) an initial collision where a parton
gas is produced; (b) a prethermal stage where elastic scatterings among partons
lead to local momentum isotropy [20]; (c) a thermal stage where parton
numbers increase until freeze-out. The term 'partonic system' refers to the
assembly
of partons in the prethermal and thermal stages. The parton plasma only
denotes the
assembly of partons in the thermal stage. The $c\bar c$ pairs are produced in
the
initial collision, prethermal and thermal stages but disintegrate in the
latter two stages. In order to understand and make
predictions for $J/\psi$ yields of RHIC and LHC experiments, the following
physical processes are taken into account. 
(a) In the initial collision, $c\bar c$ pairs are produced in hard
and semihard scatterings between partons from incoming nuclei by $2 \to 2$
processes which start at order $\alpha_s^3$ through the partonic channels $ab
\to c\bar {c}[ ^{2S+1}L_J] \rm x$. In the prethermal and thermal stages,
$c\bar c$ pairs can also be produced in $2 \to 1$ collisions which start at
order $\alpha_s^2$ via the partonic channels $ab \to
c\bar {c}[ ^{2S+1}L_J]$ since partons in the deconfined medium have large
transverse momenta.
(b) The $c\bar c$ produced at short distance is in a color singlet $(c \bar
{c})_1$ or color
octet $(c \bar {c})_8$ configuration which has a certain probability to evolve
nonperturbatively into a
color singlet state. This $J/\psi$ production process is formulated
in nonrelativistic QCD. (c) Since the color octet to singlet transition
of $(c\bar{c})_8$ takes time, gluons in the partonic medium
couple to the color octet state and destroy this transition process.
Normally, dissociation cross sections for $g+c\bar {c} \to (c\bar
{c})_8$
depend on the pair size. Expansion of the $c\bar {c}$ from a collision
point to a full $J/\psi$ size has to be taken into account.

A physical resonance formed by a $c\bar c$ pair may be one of $J/\psi$,
$\chi_{cJ}$,
$\psi'$ and others. Since the radiative transition from a higher charmonium
state to the $J/\psi$ takes a much longer time, the transition of such a
state with nonzero $p_T$ takes place
outside the partonic system. Since the Fermilab Tevatron experiments have been
able to separate direct $J/\psi$'s from those produced in radiative $\chi_{cJ}$
decays [16], in this work we assume that the
direct $J/\psi$ production can also be extracted in heavy ion measurements. If
$\chi_{cJ}$ and $\psi'$ are considered, suppression factors for $\chi_{cJ}$ and
$\psi'$ in a deconfined medium are included in prompt $J/\psi$ production. The
identification of $J/\psi$ suppression in the medium becomes
impossible for any prompt $J/\psi$ production data. Therefore,
no contributions from higher charmonium states are taken into account in
this work.

The purpose of this work is to study the dependence of the $J/\psi$ survival
probability and number distributions
produced in central Au-Au collisions at RHIC and LHC energies on the transverse
momentum and also rapidity which will be measured
in RHIC experiments [21].
The $J/\psi$
number distributions corresponding to production of $c\bar c$ in the initial
collision are
given in Section 2. Since nuclear shadowing has been shown to influence
$J/\psi$ production
in proton-nucleus collisions [22], the nuclear modification of parton
distributions
is considered. The $J/\psi$ number distributions due to $c\bar c$ production
in the prethermal and thermal stages are given in Sections 3 and 4.  Section 5
contains dissociation cross sections for gluon-$(c\bar c)_1$ and gluon-$(c\bar
c)_8$. Numerical
results for nucleon-$c\bar c$ cross sections, $J/\psi$ number distributions and
four ratios including survival probability are presented in Section 6.
Conclusions are summarized in the final section.

\vspace{0.5cm}
\leftline{\bf 2. Initial production of $c\bar c$}
\vspace{0.5cm}
Intrinsic transverse momenta of partons inside a nucleon result in the
production of $J/\psi$ with
typical momenta comparable to the QCD scale via $2 \to 1$ partonic scattering
processes [23]. Since we want to study $J/\psi$ productions with $p_T>2$ GeV,
contributions from $2 \to 1$ partonic reactions are not considered in the
initial
nucleon-nucleon collision. The effect of intrinsic transverse momentum
smearing is rather
modest for large transverse momentum $J/\psi$ data from the Tevatron [24].
Upon omission of the intrinsic
transverse momentum,
differential cross section for $J/\psi$ production in
nucleon-nucleon
collision resulting only from $2 \to 2$ partonic processes is given as
\begin{eqnarray}
\frac {d^3 \sigma}{dydy_{\rm x}dp_{\bot}} & = & 2p_{\bot} \sum_{ab{\rm x}} x_a
x_b f_{a/N}(x_a)f_{b/N}(x_b)   \nonumber  \\
&  &
[\sum_{(1)} \frac {d \sigma}{dt}(ab \to c \bar {c} [ ^{2S+1}L^{(1)}_J] {\rm x}
\to J/\psi)     \nonumber   \\
&  &
+\sum_{(8)} \frac {d \sigma}{dt}(ab \to c \bar {c} [ ^{2S+1}L^{(8)}_J] {\rm x}
\to J/\psi)]
\end{eqnarray}
where the summation $\sum_{ab{\rm x}}$ is over partons labeled by $a, b, {\rm
x}$, $\sum_{(1)}$ for all possible color-singlet states and $\sum_{(8)}$ for
all possible color-octet states. Here $\frac {d\sigma}{dt}$ denotes the
partonic
differential cross section for producing a $c\bar c[ ^{2S+1}L_J]$ and evolving
to a $J/\psi$  with
spectroscopic notation for quantum numbers and superscripts for singlet and
octet [23, 25], and $f_{a/N}$
is the parton distribution function of the species $a$ in a free nucleon. The
longitudinal
momentum fractions carried by initial partons, $x_a$ and $x_b$, are related to
rapidities of $c \bar c$ and ${\rm x}$, $y$ and
$y_{\rm x}$, by
\begin{displaymath}
x_a=\frac {1}{\sqrt s} (m_{\bot} e^{y} +p_{\bot} e^{y_{\rm x}}),~~~~~~~~~~~
x_b=\frac {1}{\sqrt s} (m_{\bot} e^{-y} +p_{\bot} e^{-y_{\rm x}})
\end{displaymath}
where $\sqrt s$,
$p_{\bot}$ and $m_{\bot}$ are the center-of-mass energy of nucleon-nucleon
collision, transverse momentum and transverse mass of the
$J/\psi$. The conditions $x_a<1$ and $x_b<1$ restrict $y_{\rm x}$ to a
region of
\begin{displaymath}
-\ln \frac {\sqrt {s}-m_{\bot}e^{-y}}{p_{\bot}} < y_{\rm x} <
\ln \frac {\sqrt {s}-m_{\bot}e^{y}}{p_{\bot}}
\end{displaymath}
These $2 \to 2$ processes at order $\alpha_s^3$, $gg \to c \bar {c} [
^{2S+1}L_J]g$, $q \bar {q} \to
c \bar {c} [ ^{2S+1}L_J]g$, $gq \to c \bar {c} [ ^{2S+1}L_J]q$ and $g\bar {q}
\to c \bar {c} [ ^{2S+1}L_J]\bar q$, start in
initial nucleus-nucleus collisions and proceed with the expansion of the heavy
pair.
While a $c \bar c$ propagates inside a
prethermal or thermal partonic system, gluons hit and excite it to continuum
states. Let $\sigma_{gc\bar {c}[1 ^3S_1^{(1)}]}$ be the cross section for
$g+(c\bar {c})[1 ^3S_1^{(1)}] \to (c\bar {c})_8$, $\sigma_{gc\bar {c}[
S^{(8)}]}$ for $g+(c\bar {c})[S^{(8)}] \to (c\bar {c})_8$ and
$\sigma_{gc\bar {c}[
P^{(8)}]}$ for $g+(c\bar {c})[P^{(8)}] \to (c\bar {c})_8$ respectively. The
cross sections are calculated in Section 5. The probability for dissociation of
a small-size $c\bar c$ into a free state relies on
the relative velocity between the gluon and $c \bar c$, $v_{rel}$, and
gluon number densities in the prethermal and thermal
stages,
$n_g(x)$ and
$n_g(\tau)$, respectively. Here the variables $x$ and $\tau$ are
individually space-time coordinates and proper time.
In the prethermal stage, parton distributions depend on the
correlation between momentum and space-time coordinates [18, 19]. The
dependence of the
gluon number density $n_g(x)$ on $x$ characterizes the partonic
system in nonequilibrium. In the thermal stage, thermal parton distributions
can be approximated by J$\rm \ddot u$ttner distributions where the temperature
and parton fugacities depend only on the proper time [9, 18, 20]. As a
consequence, the gluon number density is only a function of $\tau$.
Including $c\bar c$ suppression in the partonic system, the finally-formed
number distribution of $J/\psi$ resulting from $c\bar c$ pairs produced in the
initial central A+B collision is given by
\begin{eqnarray}
\frac {dN^{2 \to 2}_{ini}}{dyd^2p_{\bot}} & = & 2
\int^{\ln \frac {\sqrt {s}-m_{\bot}e^{y}}{p_{\bot}}}_{
-\ln \frac {\sqrt {s}-m_{\bot}e^{-y}}{p_{\bot}}}  dy_{\rm x}
\int^{R_A}_0 dr r      \nonumber   \\
& & \sum x_a f_{a/A}(x_a, m^2_{\bot}, \vec {r}) x_b f_{b/B}(x_b, m^2_{\bot},
-\vec {r})      \nonumber    \\
&  &
\{ \frac {d \sigma}{dt}(ab \to c \bar {c} [ ^{3}S^{(1)}_1] {\rm x} \to J/\psi )
\nonumber   \\
& & \exp [-\int^{\tau_{iso}}_{\tau_0} d\tau' n_g(x')
<v_{rel} \sigma_{gc\bar {c}[1 ^3S_1^{(1)}]} (k \cdot u)>_{pre} \theta (d-V_T
\Delta t)   \nonumber   \\
& & -\int^{\tau_f}_{\tau_{iso}} d\tau' n_g(\tau')
<v_{rel} \sigma_{gc\bar {c}[1 ^3S_1^{(1)}]} (k \cdot u)>_{the} \theta (d-V_T
\Delta t)]        \nonumber   \\
& & + \frac {d \sigma}{dt}(ab \to c \bar {c} [ ^{3}S^{(8)}_1] {\rm x} \to
J/\psi )   \nonumber   \\
& & \exp [-\int^{\tau_{iso}}_{\tau_0} d\tau' n_g(x')
<v_{rel} \sigma_{gc\bar {c}[S^{(8)}]} (k \cdot u)>_{pre} \theta (d-V_T \Delta
t)         \nonumber   \\
& & -\int^{\tau_f}_{\tau_{iso}} d\tau' n_g(\tau')
<v_{rel} \sigma_{gc\bar {c}[S^{(8)}]} (k \cdot u)>_{the} \theta (d-V_T \Delta
t)]        \nonumber   \\
& & + \frac {d \sigma}{dt}(ab \to c \bar {c} [ ^{1}S^{(8)}_0] {\rm x} \to
J/\psi )          \nonumber   \\
& & \exp [-\int^{\tau_{iso}}_{\tau_0} d\tau' n_g(x')
<v_{rel} \sigma_{gc\bar {c}[S^{(8)}]} (k \cdot u)>_{pre} \theta (d-V_T \Delta
t)        \nonumber   \\
& & -\int^{\tau_f}_{\tau_{iso}} d\tau' n_g(\tau')
<v_{rel} \sigma_{gc\bar {c}[S^{(8)}]} (k \cdot u)>_{the} \theta (d-V_T \Delta
t)]       \nonumber   \\
& & + \frac {d \sigma}{dt}(ab \to c \bar {c} [ ^{3}P^{(8)}_J] {\rm x} \to
J/\psi )       \nonumber   \\
& & \exp [-\int^{\tau_{iso}}_{\tau_0} d\tau' n_g(x')
<v_{rel} \sigma_{gc\bar {c}[P^{(8)}]} (k \cdot u)>_{pre} \theta (d-V_T \Delta
t)         \nonumber   \\
& & -\int^{\tau_f}_{\tau_{iso}} d\tau' n_g(\tau')
<v_{rel} \sigma_{gc\bar {c}[P^{(8)}]} (k \cdot u)>_{the} \theta (d-V_T \Delta
t)]  \}      \nonumber   \\
\end{eqnarray}
where $f_{a/A}$ is the parton distribution function of a nucleus,
\begin{equation}
f_{a/A}(x,Q^2,\vec {r})=T_A(\vec {r}) S_{a/A} (x,\vec {r}) f_{a/N}(x,Q^2)
\end{equation}
with the thickness function $T_A$ and nuclear parton shadowing factor
$S_{a/A}$. Here, $R_A$ is the nuclear radius.
The symbols $< \cdots >_{pre}$ and $< \cdots
>_{the}$ denote averages over gluon distributions in the prethermal and thermal
stages, respectively. Along the track of nucleus-nucleus collisions, a
deconfined partonic gas is produced from scatterings among primary partons at
$\tau_0$,
then reaches thermalization at $\tau_{iso}$ and finally freezes out
at
$\tau_f$. Here, $d$ is the shortest distance which a $c\bar c$ travels from a
production
point $\vec r$ to the surface of the partonic medium with transverse velocity
$V_T$ [12]. Suppose
a $c \bar c$ is produced at a proper time $\tau$ and a spatial rapidity $\eta$.
The time $\Delta t$ for the partonic system to evolve to another
proper time $\tau'$ is
\begin{equation}
\Delta t =\frac { (V_{\parallel} \sinh \eta - \cosh \eta ) \tau +
\sqrt {(\sinh \eta -V_{\parallel} \cosh \eta)^2 \tau^2
+(1-V_{\parallel}^2)\tau'^2} } {1-V_{\parallel}^2}
\end{equation}
where $V_{\parallel}$ is the longitudinal component of the $c \bar c$ velocity.
The disappearance of medium interactions on the $c\bar c$ is ensured by the
step function $\theta$ while this pair escapes from the partonic medium.

\vspace{0.5cm}
\leftline{\bf 3. Production of $c \bar {c}$ in the prethermal stage}
\vspace{0.5cm}
To order $\alpha^2_s$,
a $c \bar {c}$ in a color singlet state
is produced only through gluon fusion $gg \to c\bar {c} [
^{2S+1}L_J^{(1)}]$. For the $c \bar {c} [ ^{3}S_1^{(1)}]$, this fusion
does not occur. In contrast,
color octet states result from both channels $gg \to c \bar {c} [
^{2S+1}L_J^{(8)}]$ and $q\bar {q} \to c \bar {c}[ ^{2S+1}L_J^{(8)}]$.
Nevertheless, the number densities of quarks and antiquarks are so small that
they are neglected in estimating the production of $c\bar c$ in the prethermal
stage where
gluons dominate the partonic system.
Four momenta of the
two initial partons and final $c \bar c$ are denoted by $k_1=(\omega_1, \vec
{k}_1)$,
$k_2=(\omega_2, \vec {k}_2)$ and $p=(E, \vec {p})=(m_{\bot} \cosh y, \vec
{p}_{\bot},
m_{\bot} \sinh y)$. The differential production
rate for $gg \to c \bar {c} [ ^{2S+1}L^{(8)}_J] \to J/\psi$ in the prethermal
stage is
\begin{eqnarray}
E\frac{d^3A^{2 \to 1}_{pre}}{d^3p} & = & \frac{1}{8(2\pi)^5} \int
\frac{d^3k_1}{\omega_1}
\frac{d^3k_2}{\omega_2} \delta^{(4)}(k_1+k_2-p)
\frac{1}{2}g^2_G
f_g(k_1,x)f_g(k_2,x)              \nonumber    \\
& &  \sum_{(8)} \mid {\cal M} (gg \to c\bar{c}[ ^{2S+1}L_J^{(8)}] \to J/\psi)
\mid^2              \nonumber   \\
\end{eqnarray}
where $g_G$ is the degeneracy factor for gluons and the $f_g(k,x)$ is the
correlated phase-space distribution function given in Ref. [18].
The squared amplitudes $\mid {\cal M} \mid^2$ for $c\bar c$ in color singlet
and color octet are calculated individually in Refs. [23, 25].
To order $\alpha_s^2$, the allowed color octet states are $^1S_0^{(8)}$
and $^3P_{0,2}^{(8)}$ through the gluon fusion channel. Taking into account
the suppression of $c \bar {c}$ in
the prethermal and thermal stages, the finally-formed number distribution of
$J/\psi$ resulting
from $c\bar c$ pairs produced through $2 \to 1$ processes in the prethermal
stage is given by
\begin{eqnarray}
\frac {dN^{2 \to 1}_{pre}}{dyd^2p_{\bot}} & = &
\frac {\pi R^2_A}{16(2\pi)^5}
\int^{\tau_{iso}}_{\tau_0} \tau d\tau d\eta
d \phi_{k_1} dy_{k_1} \frac {k^2_{\bot 1}}{m^2_c}
\frac {1}{2} g^2_G  f_g(k_1,x)f_g(k_2,x)    \nonumber    \\
& & \{ \mid {\cal M} (gg \rightarrow c\bar{c} [ ^{1}S_0^{(8)}] \to
J/\psi) \mid^2       \nonumber   \\
& & \exp [-\int^{\tau_{iso}}_{\tau} d\tau' n_g(x')
<v_{rel} \sigma_{gc\bar {c}[S^{(8)}]} (k \cdot u)>_{pre} \theta (d-V_T \Delta
t) \nonumber \\
& & -\int^{\tau_f}_{\tau_{iso}} d\tau' n_g(\tau')
<v_{rel} \sigma_{gc\bar {c}[S^{(8)}]} (k \cdot u)>_{the} \theta (d-V_T \Delta
t)] \nonumber   \\
& & +  \mid {\cal M} (gg \rightarrow c\bar{c} [ ^{3}P_J^{(8)}] \to
J/\psi) \mid^2       \nonumber   \\
& & \exp [-\int^{\tau_{iso}}_{\tau} d\tau' n_g(x')
<v_{rel} \sigma_{gc\bar {c}[P^{(8)}]} (k \cdot u)>_{pre} \theta (d-V_T \Delta
t) \nonumber \\
& & -\int^{\tau_f}_{\tau_{iso}} d\tau' n_g(\tau')
<v_{rel} \sigma_{gc\bar {c}[P^{(8)}]} (k \cdot u)>_{the} \theta (d-V_T \Delta
t)]  \} \nonumber    \\
\end{eqnarray}
where
$\phi_{k_i}$ is the angle between $\vec {k}_{\bot i}$ and $\vec {p}_{\bot}$
for $i=1,2$ and
$m_c$ is the charm quark mass.
The kinematic variables $k_{\bot 1}$, $k_{\bot 2}$, $\phi_{k_2}$ and
$y_{k_2}$ are expressed in terms of
\begin{displaymath}
k_{\bot 1}=\frac {2m^2_c} {m_{\bot} \cosh(y-y_{k_1})-p_{\bot} \cos
\phi_{k_1}}
\end{displaymath}
\begin{displaymath}
k_{\bot 2}=\sqrt {p_{\bot}^2 +k_{\bot 1}^2 -2 p_{\bot} k_{\bot 1} \cos
\phi_{k_1} }
\end{displaymath}
\begin{displaymath}
\sin \phi_{k_2} = -\frac {k_{\bot 1}}{k_{\bot 2}} \sin \phi_{k_1}
\end{displaymath}
\begin{displaymath}
\sinh y_{k_2} = \frac {1}{k_{\bot 2}}(m_{\bot} \sinh y -k_{\bot 1} \sinh
y_{k_1})
\end{displaymath}

To order $\alpha_s^3$,
the differential production rate gets contributions from the processes
$gg \to c \bar {c} [ ^{2S+1}L_J^{(1,8)}]g$ in the prethermal stage,
\begin{eqnarray}
E \frac {d^3A^{2 \to 2}_{pre}}{d^3p} & = & \frac {1}{16(2\pi)^8} \int \frac
{d^3k_1}{\omega_1} \frac {d^3k_2}{\omega_2} \frac {d^3p_{\rm x}}{E_{\rm x}}
\delta^{(4)} (k_1+k_2-p-p_{\rm x})       \nonumber    \\
& &  \frac {1}{2} g^2_G f_g(k_1,x)f_g(k_2,x)
\{ \sum_{(1)} \mid {\cal {M}} (gg \to c\bar {c}[ ^{2S+1}L_J^{(1)}]{\rm x} \to
J/\psi) \mid^2                 \nonumber    \\
& & + \sum_{(8)} \mid {\cal {M}} (gg \to c\bar {c}[ ^{2S+1}L_J^{(8)}]{\rm x}
\to J/\psi) \mid^2  \}
\end{eqnarray}
where $p_{\rm x}=(E_{\rm x},\vec {p}_{\rm x})=(p_{\bot {\rm x}}\cosh y_{\rm x},
{p}_{\bot {\rm x}} \cos \phi_{\rm x}, {p}_{\bot {\rm x}} \sin \phi_{\rm x},
p_{\bot {\rm x}} \sinh y_{\rm x})$ is the four momentum of the massless parton
{\rm x}.
Taking into account the suppression of $c \bar {c}$ in the
prethermal and thermal stages, the finally-formed number distribution  of
$J/\psi$ resulting
from $c\bar c$ pairs produced through $2 \to 2$ processes in the prethermal
stage is given by
\begin{eqnarray}
\frac {dN^{2 \to 2}_{pre}}{dyd^2p_{\bot}}
& = & \frac {\pi R_A^2}{16(2\pi)^8}
 \int^{\tau_{iso}}_{\tau_0} \tau d\tau d\eta
 p_{\bot {\rm x}} dp_{\bot {\rm x}} d\phi_{\rm x} dy_{\rm x} d\phi_{k_1}
  dy_{k_1}     \nonumber   \\
& & \frac {2k^2_{\bot 1}}{\hat s} \frac {1}{2} g^2_G f_g(k_1,x)f_g(k_2,x)
\nonumber  \\
& & \{ \mid {\cal {M}} (gg \to c\bar {c}[ ^{3}S_1^{(1)}]{\rm x} \to J/\psi)
\mid^2            \nonumber  \\
& & \exp [-\int^{\tau_{iso}}_{\tau} d\tau' n_g(x')
<v_{rel} \sigma_{gc\bar {c}[1 ^3S_1^{(1)}]} (k \cdot u)>_{pre} \theta (d-V_T
\Delta t) \nonumber   \\
& & -\int^{\tau_f}_{\tau_{iso}} d\tau' n_g(\tau')
<v_{rel} \sigma_{gc\bar {c}[1 ^3S_1^{(1)}]} (k \cdot u)>_{the} \theta (d-V_T
\Delta t)] \nonumber   \\
& & + \mid {\cal M} (gg \to c \bar {c} [ ^{3}S^{(8)}_1] {\rm x} \to
J/\psi ) \mid^2    \nonumber   \\
& & \exp [-\int^{\tau_{iso}}_{\tau} d\tau' n_g(x')
<v_{rel} \sigma_{gc\bar {c}[S^{(8)}]} (k \cdot u)>_{pre} \theta (d-V_T \Delta
t) \nonumber   \\
& & -\int^{\tau_f}_{\tau_{iso}} d\tau' n_g(\tau')
<v_{rel} \sigma_{gc\bar {c}[S^{(8)}]} (k \cdot u)>_{the} \theta (d-V_T \Delta
t)] \nonumber   \\
& & + \mid {\cal M} (gg \to c \bar {c} [ ^{1}S^{(8)}_0] {\rm x} \to
J/\psi )  \mid^2        \nonumber   \\
& & \exp [-\int^{\tau_{iso}}_{\tau} d\tau' n_g(x')
<v_{rel} \sigma_{gc\bar c[S^{(8)}]} (k \cdot u)>_{pre} \theta (d-V_T \Delta t)
\nonumber   \\
& & -\int^{\tau_f}_{\tau_{iso}} d\tau' n_g(\tau')
<v_{rel} \sigma_{gc\bar c[S^{(8)}]} (k \cdot u)>_{the} \theta (d-V_T \Delta t)]
\nonumber   \\
& & + \mid {\cal M} (gg \to c \bar {c} [ ^{3}P^{(8)}_J] {\rm x} \to
J/\psi )   \mid^2    \nonumber   \\
& & \exp [-\int^{\tau_{iso}}_{\tau} d\tau' n_g(x')
<v_{rel} \sigma_{gc\bar {c}[P^{(8)}]} (k \cdot u)>_{pre} \theta (d-V_T \Delta
t) \nonumber   \\
& & -\int^{\tau_f}_{\tau_{iso}} d\tau' n_g(\tau')
<v_{rel} \sigma_{gc\bar {c}[P^{(8)}]} (k \cdot u)>_{the} \theta (d-V_T \Delta
t)]  \} \nonumber   \\
\end{eqnarray}
where $\hat {s} =(k_1+k_2)^2$ and some kinematic variables are given by
\begin{displaymath}
k_{\bot 1}  = \{ 4m_c^2+2m_{\bot} p_{\bot {\rm x}} \cosh(y-y_{\rm x}) -2
p_{\bot}p_{\bot {\rm x}}
\cos \phi_{\rm x} \} /
\end{displaymath}
\begin{displaymath}
\{ 2[m_{\bot} \cosh(y-y_{k_1}) +p_{\bot {\rm
x}}\cosh (y_{\rm x}-y_{k_1}) -p_{\bot} \cos \phi_{k_1} -p_{\bot {\rm x}}\cos
(\phi_{\rm x} -\phi_{k_1})] \}
\end{displaymath}
\begin{eqnarray}
k^2_{\bot 2} & = & m^2_{\bot} +p^2_{\bot {\rm x}} +2 m_{\bot} p_{\bot
{\rm x}} \cosh(y-y_{\rm x}) +k^2_{\bot 1}     \nonumber   \\
&  &  -2k_{\bot 1} [m_{\bot} \cosh(y-y_{k_1}) +p_{\bot {\rm x}}
\cosh (y_{\rm x}-y_{k_1})]     \nonumber
\end{eqnarray}
\begin{displaymath}
\sinh y_{k_2} =\frac {1}{k_{\bot {\rm x}}} [m_{\bot} \sinh y +p_{\bot
{\rm x}} \sinh y_{\rm x} -k_{\bot 1} \sinh y_{k_1}]
\end{displaymath}

The $J/\psi$ number distribution resulting from $c\bar c$ pairs produced in the
prethermal stage becomes
\begin{equation}
\frac {dN_{pre}}{dyd^2p_{\bot}}  =
\frac {dN^{2 \to 1}_{pre}}{dyd^2p_{\bot}} +
\frac {dN^{2 \to 2}_{pre}}{dyd^2p_{\bot}}
\end{equation}

\vspace{0.5cm}
\leftline{\bf 4. Production of $c \bar c$ in the thermal stage}
\vspace{0.5cm}
In the thermal stage, parton distributions are approximated by thermal
phase-space distributions $f_i(k;T,\lambda_i)$ in which the temperature $T$ and
nonequilibrium
fugacities $\lambda_i$ are functions of the proper time $\tau$ [9, 18]. While
the
partonic system evolves, quark and antiquark number densities increase. To
order $\alpha_s^2$, both $gg \to c \bar {c}[ ^{2S+1}L_J^{(8)}] \to J/\psi$
and $q\bar {q} \to c \bar {c}[ ^{2S+1}L_J^{(8)}] \to J/\psi$ contribute to
the $J/\psi$ number distribution in the thermal stage
\begin{eqnarray}
\frac {dN^{2 \to 1}_{the}}{dyd^2p_{\bot}} & = &
\frac {\pi R^2_A}{16(2\pi)^5}
\int^{\tau_f}_{\tau_{iso}} \tau d\tau d\eta
d \phi_{k_1} dy_{k_1} \frac {k^2_{\bot 1}}{m^2_c}     \nonumber    \\
&  & \{ \frac {1}{2} g^2_G f_g(k_1;T,\lambda_g)f_g(k_2;T,\lambda_g)
\mid {\cal M} (gg \to c\bar{c} [ ^{1}S_0^{(8)}] \to J/\psi) \mid^2
\nonumber \\
& & \exp [-\int^{\tau_f}_{\tau} d\tau' n_g(\tau')
<v_{rel} \sigma_{gc\bar {c}[S^{(8)}]} (k \cdot u)>_{the} \theta (d-V_T \Delta
t) ] \nonumber    \\
&  & + \frac {1}{2} g^2_G f_g(k_1;T,\lambda_g)f_g(k_2;T,\lambda_g)
\mid {\cal M} (gg \to c\bar{c} [ ^{3}P_J^{(8)}] \to J/\psi) \mid^2
\nonumber \\
& & \exp [-\int^{\tau_f}_{\tau} d\tau' n_g(\tau')
<v_{rel} \sigma_{gc\bar {c}[P^{(8)}]} (k \cdot u)>_{the} \theta (d-V_T \Delta
t) ] \nonumber    \\
&  & +g_q g_{\bar {q}} f_q(k_1;T,\lambda_q) f_{\bar {q}}(k_2;T,\lambda_{\bar
q}) \mid {\cal M}
(q\bar {q} \to c\bar{c} [ ^{3}S_1^{(8)}] \to J/\psi ) \mid^2  \nonumber  \\
& & \exp [-\int^{\tau_f}_{\tau} d\tau' n_g(\tau')
<v_{rel} \sigma_{gc\bar {c}[S^{(8)}]} (k \cdot u)>_{the} \theta (d-V_T \Delta
t) ] \} \nonumber   \\
\end{eqnarray}
where $g_q$ and $g_{\bar q}$ are the degeneracy factors for quarks and
antiquarks, respectively. In the channel of quark-antiquark annihilation, only
the squared amplitude for $^3S_1^{(8)}$ does not vanish.

All lowest-order $2 \to 2$ reactions $gg \to c \bar {c} [ ^{2S+1}L_J]g$, $q
\bar {q} \to
c \bar {c} [ ^{2S+1}L_J]g$, $gq \to c \bar {c} [ ^{2S+1}L_J]q$ and $g\bar {q}
\to c \bar {c} [ ^{2S+1}L_J]\bar q$ contribute to the
 $J/\psi$ number distribution in the thermal stage
\begin{eqnarray}
\frac {dN^{2 \to 2}_{the}}{dyd^2p_{\bot}}
& = & \frac {\pi R_A^2}{16(2\pi)^8}
 \int^{\tau_f}_{\tau_{iso}} \tau d\tau d\eta
 p_{\bot {\rm x}} dp_{\bot {\rm x}} d\phi_{\rm x} dy_{\rm x} d\phi_{k_1}
dy_{k_1}  \frac {2k^2_{\bot 1}}{\hat s}        \nonumber   \\
& &  \sum_{ab} \frac {1}{2}  f_a(k_1;T,\lambda_a)f_b(k_2;T,\lambda_b) g_a g_b
\nonumber  \\
& & \{ \sum_{(1)} \mid {\cal {M}} (ab \to c\bar {c}[ ^{2S+1}L_J^{(1)}]{\rm x}
\to J/\psi) \mid^2                 \nonumber    \\
& & \exp [-\int^{\tau_f}_{\tau} d\tau' n_g(\tau')
<v_{rel} \sigma_{gc\bar {c}[1 ^{2S+1}L_J^{(1)}]} (k \cdot u)>_{the} \theta
(d-V_T \Delta t)] \nonumber   \\
& & +  \sum_{(8)} \mid {\cal {M}} (ab \to c\bar {c}[ ^{2S+1}L_J^{(8)}]{\rm x}
\to J/\psi) \mid^2            \nonumber  \\
& & \exp [-\int^{\tau_f}_{\tau} d\tau' n_g(\tau')
<v_{rel} \sigma_{gc\bar {c}[L^{(8)}]} (k \cdot u)>_{the} \theta
(d-V_T \Delta t)]  \} \nonumber   \\
\end{eqnarray}

The $J/\psi$ number distribution resulting from $c\bar c$ pairs produced in the
thermal stage becomes
\begin{equation}
\frac {dN_{the}}{dyd^2p_{\bot}}  =
\frac {dN^{2 \to 1}_{the}}{dyd^2p_{\bot}} +
\frac {dN^{2 \to 2}_{the}}{dyd^2p_{\bot}}
\end{equation}

\vspace{0.5cm}
\leftline{\bf 5. Gluon-$c\bar {c}$ dissociation cross sections}
\vspace{0.5cm}
A dissociation cross section of a full-size $J/\psi$ induced by a gluon is
given in
Refs. [1, 26]. Since an initially-created $c \bar c$ has a radius of about
$r_0=\frac {1}{2m_c}$ and proceeds by expanding to a full-size object, the
dissociation cross section of
$c\bar c$ by a gluon has a size dependence. By this we mean the dissociation of
$c\bar c$ into free states via this process $g+c\bar {c} \to (c\bar {c})_8$.
Cross sections are calculated with
chromoelectric dipole coupling between gluon and $c \bar c$
in the procedure for gluon-$J/\psi$ dissociation in Ref. [26]. The
wave function of an expanding $c\bar c$ is needed for this purpose, but it
has not been investigated in the partonic medium even though some
attempts have been made in studies of the color transparency phenomenon [27].
We proceed with the construction of wave functions in a simple
one-gluon-exchange potential model.

In a parton plasma, the internal motion of $J/\psi$ is obtained [12] from the
attractive Coulomb
potential, $V_1=-{\rm {g}_s}^2/3\pi r$. The quantum-mechanical interpretation
of the $c \bar c$ radius is $\sqrt {<r^2>}$,
the square root of the radius-square expectation value of
the relative-motion wave function. For the 1S color singlet, its wave function
in momentum space
normalized to the radius of $c \bar c [1 ^{3}S_1^{(1)}]$ is
\begin{eqnarray}
[\vec {r} \psi_{1s}](\vec {k})=32 \sqrt {\pi} a_0^{2.5} \frac {\vec {k} a_0}
{(1+(ka_0)^2)^3}
\end{eqnarray}
where the variable $a_0=\sqrt {<r^2>/3}$ is the Bohr radius for a full-size
$J/\psi$. The velocity-square expectation value of
the $J/\psi$ wave function is $<v^2>=0.428$. Then the radius of $c \bar c$ is
assumed to
expand according to $\sqrt {<r^2>}=\sqrt {<v^2>} t + r_0$.
The gluon-$c \bar {c} [1 ^{3}S_1^{(1)}]$ dissociation cross section is
\begin{equation}
\sigma_{gc\bar {c} [1 ^3S_1^{(1)}]}
=\frac
{128{\rm
{g}_s}^2m_c^{2.5}a_0^7(Q^0-\epsilon_0)^{1.5}Q^0}{9[m_ca_0^2(Q^0-\epsilon_0)+1]
^ 6 }
\end{equation}
where $Q^0$ is the gluon energy, $\epsilon_0$ the binding energy of
$J/\psi$ and $\rm {g}_s$ the strong coupling constant.

While the $c\bar c$ is in a color octet state, it is not a bound state but
rather a scattering state. Its relative-motion wave function
is determined by the repulsive potential $V_8={\rm {g}_s^2}/{24\pi r}$.
The radial part of the $S$ wave function is
\begin{equation}
S_R(r)+{\rm i}S_I(r)=e^{{\rm i}qr}F(1+{\rm i}\eta,2,-2{\rm i}qr)
\end{equation}
and the radial part of the $P$ wave function is
\begin{equation}
P_R(r)+{\rm i}P_I(r)=qre^{{\rm i}qr}F(2+{\rm i}\eta,4,-2{\rm i}qr)
\end{equation}
with $q=m_c \sqrt {<v^2>}$ and $\eta={\rm {g}_s^2}/{24\pi\sqrt {<v^2>}}$.
The function $F$ is the confluent hypergeometric function.
Wave functions in momentum space are obtained by performing a Fourier
transform of the wave functions in space coordinates. Normalization
constants of the momentum-space wave functions, $C_S$ and $C_P$, are determined
by fitting the $c\bar c$ radius.
Dissociation cross sections of the $S$-wave and $P$-wave color-octet states
by a gluon are
\begin{eqnarray}
\sigma_{gc\bar {c} [S^{(8)}]} & = &
\frac {{\rm {g}_s^2}  (m_cQ^0)^{1.5}}{576\pi}C^2_S
\int^b_0 \int^b_0 dr_1dr_2 r^3_1r^3_2    \nonumber   \\
& & j_1(\sqrt {m_cQ^0}r_1) j_1(\sqrt {m_cQ^0}r_2)
[S_R(r_1)S_R(r_2)+S_I(r_1)S_I(r_2)]    \nonumber  \\
\end{eqnarray}
\begin{eqnarray}
\sigma_{gc\bar {c} [P^{(8)}]} & = &
\frac {{\rm {g}_s^2} (m_cQ^0)^{1.5}}{576\pi}C^2_P
\int^b_0 \int^b_0 dr_1dr_2 \frac {r^3_1r^3_2}{6\pi^2}
\nonumber   \\
& & [j_0(\sqrt {m_cQ^0}r_1) j_0(\sqrt {m_cQ^0}r_2)
    +2j_2(\sqrt {m_cQ^0}r_1) j_2(\sqrt {m_cQ^0}r_2)]    \nonumber   \\
& & [P_R(r_1)P_R(r_2)+P_I(r_1)P_I(r_2)]    \nonumber  \\
\end{eqnarray}
where the $j_0$, $j_1$ and $j_2$ are spherical Bessel functions.
The $b$ is determined so that the square root of the $r^2$ expectation value
of the relative wave function in Eq. (15) or (16) is the color-octet radius.
Relations $b=1.435 \sqrt {<r^2>}$ for $S$-wave and $b=1.3 \sqrt {<r^2>}$ for
$P$-wave approximately hold for color-octet size less than normal hadron
size.

\vspace{0.5cm}
\leftline{\bf 6. Numerical results and discussions}
\vspace{0.5cm}
Results for five aspects are presented in the following subsections. The first
aspect is the nucleon-$c\bar c$ dissociation cross sections shown in the next
subsection. The second one in Subsection 6.2 is $J/\psi$ number distributions
versus transverse momentum at $y=0$ and rapidity at $p_T=4$ GeV with nuclear
effect on parton distributions  and $c\bar c$ dissociation in the partonic
system. The third one in Subsection 6.3 is to define and calculate four ratios
including survival probability with $y=0$ or
$p_T=4$ GeV at both RHIC and LHC energies. The fourth one is given in
Subsection 6.4
to show $J/\psi$ number distributions without nuclear effect on parton
distributions
and $c\bar c$ dissociation in the partonic system. The fifth one concerns some
uncertainties on the above results.

\vspace{0.5cm}
\leftline{\it 6.1. Nucleon-$c\bar c$ dissociation cross sections}
\vspace{0.5cm}
In the parton model of the nucleon, the gluon is a dominant ingredient.
Whereas the
cross section for $c\bar c$ dissociated directly by a real gluon is of order
$\alpha_s$, the cross section for the quark-$c\bar c$ dissociation through a
virtual
gluon is of order $\alpha_s^2$. With the gluon-$(c\bar {c})_1$ cross section
given
in the last section, the nucleon-$c\bar {c}[1 ^3S_1^{(1)}]$ cross section
driven mainly by the gluon ingredient becomes
\begin{equation}
\sigma_{Nc\bar {c}[1 ^3S_1^{(1)}]}
       =\int^1_{x^{(1)}_{min}} dx f_{g/N}(x,Q^2)
\sigma_{gc\bar {c}[1 ^3S_1^{(1)}]}
\end{equation}
where
$x_{min}^{(1)}=\frac {\epsilon_0}{p_N}$ with
$p_N$ being the proton momentum in  the rest frame of the $J/\psi$.
The gluon distribution function $f_{g/N}$ is
that Gl$\rm \ddot {u}$ck-Reya-Vogt (GRV) result at leading order in Ref. [28].
The cross section is drawn in Fig. 1 to show the energy and
renormalization-scale dependence while the $c\bar c$ radius is the
$J/\psi$ radius in the attractive Coulomb potential, $r_{J/\psi}=0.348$ fm.
In Fig. 1, gluon field
operators are renormalized at three scales $\epsilon_0, \sqrt {2}\epsilon_0,
Q^0$, respectively. The coupling constant has the value
$\alpha_s=\frac {4}{3}\sqrt {\frac {\epsilon_0}{m_c}}$ corresponding to the
scale $\epsilon_0$ while it varies for the other two scales.
Values
of the cross section at $\sqrt {s}=10$ GeV are a little lower than the
nucleon-$J/\psi$ dissociation cross section obtained by
the subtraction of quasi-elastic cross section in Ref. [29] from the total
cross section given in Ref. [30].
In high-temperature hadronic matter or $J/\psi$ photoproduction reaction,
a typical value of the center-of-mass energy for nucleon-$J/\psi$ (or
preresonance) dissociation is around $\sqrt
{s}=6$ GeV [1]. At this energy, Fig. 2 is drawn to show the size dependence of
$\sigma_{Nc\bar {c}[1 ^3S_1^{(1)}]}$,
with $f_{g/N}(x,Q^2)$ at $Q^2=\epsilon_0^2$.

For the $S$-wave color octet the
nucleon-$c\bar {c}[S^{(8)}]$ cross section is
\begin{equation}
\sigma_{Nc\bar {c}[S^{(8)}]}=\int^1_{x^{(8)}_{min}} dx f_{g/N}(x,
(q+ Q^0)^2) \sigma_{gc\bar {c}[S^{(8)}]}
\end{equation}
where $x_{min}^{(8)}=\frac {\Lambda_{QCD}}{p_N}$.
For the $P$-wave color octet the
nucleon-$c\bar {c}[P^{(8)}]$ cross section is
\begin{equation}
\sigma_{Nc\bar {c}[P^{(8)}]}=\int^1_{x^{(8)}_{min}} dx f_{g/N}(x,
(q+ Q^0)^2) \sigma_{gc\bar {c}[P^{(8)}]}
\end{equation}
Since the gluon momentum in a confining medium is
bigger than the QCD scale [14], the lowest value of $x$ is set by
$\Lambda_{QCD}=0.2322$ GeV used in the leading order GRV parton
distribution functions. Dependences of
$\sigma_{Nc\bar {c}[S^{(8)}]}$ and
$\sigma_{Nc\bar {c}[P^{(8)}]}$
on the center-of-mass energy $\sqrt {s}$ are depicted in Fig. 3 while the size
of $(c\bar {c})_8$ is the full size of $J/\psi$. The dot-dashed line is
obtained with the nucleon-$J/\psi$ cross section given by Eq. (24) in Ref. [1]
where another gluon distribution
function evaluated at $Q^2=\epsilon_0^2$ is used.
While the $(c\bar
c)_8$ has small momentum in a nucleus, the cross section for nucleon-$(c \bar
{c})_8$ production is lower than the absorption cross section determined by
Gerschel and H$\rm \ddot u$fner [31]
or the two-gluon exchange result [32].
It was proposed by Kharzeev and Satz that the color octet plus a gluon
configuration is a dominant component produced in
the proton-nucleus collisions [14]. In fact, the present cross section for
a nucleon and a bare $(c\bar {c})_8$ is one part of
the nucleon-$g(c\bar {c})_8$ cross
section.

A $(c\bar {c})_8$ pair produced at a collision point expands before
becoming color singlet to a size which may be
larger or smaller than the full size of $J/\psi$. We then show in Fig. 4 the
dependence of the nucleon-$(c\bar {c})_8$ cross section on the color-octet pair
radius.

We do not want to address proton-nucleus collisions in terms of
nucleon-$c\bar c$ cross sections [33, 34] since only the
gluon-$c\bar c$ cross sections are needed to study $c\bar c$ suppression in
the prethermal and thermal stages.
In proton-nucleus collisions,
once a color-octet $(c\bar {c})_8$ pair is produced, it picks up a collinear
gluon to form a colorless configuration [14]. However,
in central Au-Au
collisions at RHIC and LHC energies, the accompanying gluon scatters with
other hard gluons in the dense partonic system and is driven away. Therefore
the bare $c\bar c$ is the object that we want to study in the partonic system.

\vspace{0.5cm}
\leftline{\it 6.2. $J/\psi$ number distributions with suppression}
\vspace{0.5cm}
$J/\psi$ number distributions versus transverse momentum  at $y=0$ and rapidity
at $p_T=4$ GeV for
central Au-Au collisions at RHIC energy $\sqrt
{s}=200A$GeV are calculated with respect to the initial collision, prethermal
and thermal stages. Initial productions of $c\bar c$ are calculated with GRV
parton distribution functions at renormalization scale $\mu=
\sqrt {p_{\bot}^2+4m_c^2}$. Evolution of color-octet states $ ^{3}S_1$, $
^{1}S_0$
and $ ^{3}P_J$ toward the $J/\psi$ is specified by nonperturbative matrix
elements $<{\cal O}^{J/\psi}_8( ^3S_1)>$, $<{\cal O}^{J/\psi}_8( ^1S_0)>$ and
$<{\cal O}^{J/\psi}_8( ^3P_0)>$ in nonrelativistic QCD [17].
In the nonperturbative evolution, a gluon from the partonic system hits and
prevents the color octet from color neutralizing via
$g+(c\bar {c})_8 \to (c\bar
{c})_8$. This medium effect has been expressed by exponentials in Eqs. (2),
(6), (8), (10) and (11).
Therefore, the nonperturbative matrix elements
are assumed to be invariant while the medium effect is factorized into
exponential forms. Values of these  matrix elements are well
determined by fitting the CDF measurements for  $p\bar p$ collisions at $\sqrt
{s}=1.8$ TeV in Ref. [35],
\begin{equation}
<{\cal O}^{J/\psi}_8( ^3S_1)>  =  (1.12 \pm 0.14)\times 10^{-2}{\rm GeV}^3
\end{equation}
\begin{equation}
<{\cal O}^{J/\psi}_8( ^1S_0)> + \frac {3.5}{m_c^2}
<{\cal O}^{J/\psi}_8( ^3P_0)>  =  (3.90 \pm 1.14) \times 10^{-2}{\rm GeV}^3
\end{equation}
In $p\bar p$ collisions, differential
cross sections of direct $J/\psi$ production depend on the combination of
$<{\cal O}^{J/\psi}_8( ^1S_0)>$ and
$<{\cal O}^{J/\psi}_8( ^3P_0)>$. However, since the dissociation cross
section for the $S$-wave color-octet state is different from that for the
$P$-wave color-octet state, such a
dependence on the combination is destroyed. In calculations, values are taken
as follows,
\begin{equation}
<{\cal O}^{J/\psi}_8( ^1S_0)> = 4 \times 10^{-2}{\rm GeV}^3,~~
<{\cal O}^{J/\psi}_8( ^3P_0)> = -\frac {m^2_c}{35} \times 10^{-2}{\rm GeV}^3
\end{equation}
The value of $<{\cal O}^{J/\psi}_8( ^3P_0)>$ is positive at tree level and
negative
after renormalization [36]. Eq. (23) is still satisfied by the values in
Eq. (24). These values of nonperturbative matrix elements are supposed to be
universal for any center-of-mass energy $\sqrt {s}$.

Various contributions to the $J/\psi$ number distributions including initial
collisions, prethermal and thermal stages, $2 \to 1$ and $2 \to 2$ collisions,
are drawn separately in Figs. 5 and 6. The dashed curve resulting from $c\bar
c$ production in the initial collision is obtained by calculating Eq. (2) where
the nuclear parton shadowing factor is given in Ref. [4] throughout this
subsection. The upper and lower dot-dashed curves resulting from $c\bar c$
production in the prethermal stage are obtained by individually calculating Eq.
(6) for $2 \to 1$ collisions and Eq. (8) for $2 \to 2$ collisions. The upper
and lower dotted curves resulting from $c\bar c$ production in the thermal
stage are obtained by calculating Eq. (10) for $2 \to 1$ collisions and Eq.
(11) for $2 \to 2$ collisions, respectively.  To exclude the
effect of intrinsic transverse momentum smearing, only the region $p_T >2$ GeV
is considered. Consequently,
no $2 \to 1$ collisions contribute in the initial collision. The
$J/\psi$
number distribution resulting from the initial collision shown by the dashed
line has
a plateau similar to that in proton-proton collision [37].
Both Figs. 5 and 6 show that $c\bar {c}$ pairs produced
from the thermal stage can be neglected compared to the initial production, but
the contributions from the prethermal stage are important in the
transverse momentum region $2 {\rm GeV}<p_T<8 {\rm GeV}$ 
and rapidity region $0<y<1.2$.
Productions of $c\bar c$ in the prethermal and thermal stages
bulge up the $J/\psi$ number distribution shown by the solid line in this
rapidity region. Nevertheless, the
dot-dashed and dotted lines fall rapidly as the rapidity gets large.
This bulging
characterizes the formation of a deconfined medium because the medium
has average momentum limited but big
enough to produce extra $c\bar c$ and  thus $J/\psi$. The
$2 \to 1$ collisions in the partonic system have bigger contributions than the
$2 \to 2$ collisions.

Each of Figs. 7 and 8 contains two sets of lines to show contributions from the
color-singlet and color-octet pairs produced at short distance. Any set has a
dashed line obtained from Eq. (2) for the initial collision, a dot-dashed line
from Eq. (9) for
the prethermal stage and a dotted line from Eq. (12)
for the thermal stage,
respectively. A line in the upper (lower) set for the color-octet
(color-singlet) contributions stems from the terms for $(c\bar {c})_8$ ($(c\bar
{c})_1$) states. The color octet states dominate productions
of $J/\psi$ at RHIC energy. However, the ratio of color-octet to color-singlet
contributions shown by the two solid lines at $p_T=6$ GeV is reduced from about
70 at CDF collider energy $\sqrt
{s}=1.8$ TeV to about 40 at RHIC energy. Both
contributions of color-singlet and color-octet states have similar
dependence on transverse momentum and rapidity.

Figs. 9 and 10 show transverse momentum and rapidity dependence of $J/\psi$
number distributions for central Au-Au collisions at LHC energy $\sqrt
{s}=5.5A$TeV. A prominent feature is that the $J/\psi$ number produced from the
thermal stage is comparable to that from the prethermal stage.
Compared to the initial production, $c\bar c$ and $J/\psi$ produced through
$2 \to 2$ reactions may be neglected. A bulge is observed on the plateau
in the rapidity region $0<y<1.5$. Such a
bulge can be taken as a signature for the existence of a parton plasma
at the LHC energy. Figs. 11 and 12 depict contributions from the color
singlet and color octet at LHC energy. The ratio of color-octet to
color-singlet
contributions shown by the two solid lines at $p_T=6$GeV reaches about 150.
This indicates
the color octet states become more important with the increase of $\sqrt {s}$.

\vspace{0.5cm}
\leftline{\it 6.3. Ratios including $J/\psi$ survival probability}
\vspace{0.5cm}
Nuclear shadowing results in a
modification of gluon distribution functions inside a nucleus [38] and such a
nuclear effect is represented by the shadowing factor $S_{a/A}$ in Eq. (3). If
the $S_{a/A}=1$ for no shadowing, the $dN^{2 \to 2}_{ini}/dyd^2p_{\bot}$ is
proportional to the product of atomic masses of the two colliding nuclei. If
$S_{a/A} \ne 1$ and depends on the longitudinal momentum fraction $x$, the
production of $c\bar c$ is reduced in the shadowing region and enhanced for the
anti-shadowing region. Irrespective of interactions of $ J/\psi$ with the
partonic system, $J/\psi$ number distributions produced in the initial
central A+B collision is obtained by putting all exponentials equal to 1 in
Eq. (2),
\begin{eqnarray}
\frac {dN^{2 \to 2}_{0}}{dyd^2p_{\bot}}(S_{a/A}) & = & 2
\int^{\ln \frac {\sqrt {s}-m_{\bot}e^{y}}{p_{\bot}}}_{
-\ln \frac {\sqrt {s}-m_{\bot}e^{-y}}{p_{\bot}}}  dy_{\rm x}
\int^{R_A}_0 dr r      \nonumber   \\
& & \sum x_a f_{a/A}(x_a, m^2_{\bot}, \vec {r}) x_b f_{b/B}(x_b, m^2_{\bot},
-\vec {r})      \nonumber    \\
&  &
\{ \frac {d \sigma}{dt}(ab \to c \bar {c} [ ^{3}S^{(1)}_1] {\rm x} \to J/\psi )
\nonumber   \\
& & + \frac {d \sigma}{dt}(ab \to c \bar {c} [ ^{3}S^{(8)}_1] {\rm x} \to
J/\psi )   \nonumber   \\
& & + \frac {d \sigma}{dt}(ab \to c \bar {c} [ ^{1}S^{(8)}_0] {\rm x} \to
J/\psi )
\nonumber  \\
& & + \frac {d \sigma}{dt}(ab \to c \bar {c} [ ^{3}P^{(8)}_J] {\rm x} \to
J/\psi ) \}    \nonumber   \\
\end{eqnarray}
To characterize the influence of nuclear parton shadowing on the $J/\psi$
production from the initial collision, a ratio is defined as
\begin{equation}
R^{ini}=
\frac {dN^{2 \to 2}_{0}}{dyd^2p_{\bot}}(S_{a/A} \ne 1) /
\frac {dN^{2 \to 2}_{0}}{dyd^2p_{\bot}}(S_{a/A}=1)
\end{equation}
Here the $S_{a/A} \ne 1$ from Ref. [4] applies throughout this subsection.

The initially produced $J/\psi$
originates from the $c \bar c$ pairs produced in the initial collision. Its
dependence on the transverse momentum and rapidity is obtained by calculating
Eq. (25).
Some $c\bar c$ pairs produced in the initial collision may dissociate by
gluons from the partonic system. As a consequence, the $J/\psi$ number is
reduced.
The survival probability for the $c\bar c$ transiting into a $J/\psi$ is
defined as the ratio
\begin{equation}
S^{plasma}  =  \frac {dN^{2 \to 2}_{ini}}{dyd^2p_{\bot}} (S_{a/A}\ne 1) /
      \frac {dN^{2 \to 2}_0}{dyd^2p_{\bot}} (S_{a/A}\ne 1)
\end{equation}

We have calculated the $J/\psi$ number distributions produced in the prethermal
and thermal stages in Subsection 6.2. The $J/\psi$ yield may be bigger than the
reduced amount of initially produced $J/\psi$
due to the $c \bar c$ dissociation by gluons in the partonic system.
The partonic system has two roles. One is to produce $c\bar c$ pairs and
another is to dissociate $c\bar c$ pairs.
To see the roles, a ratio is defined by
\begin{equation}
R^{plasma}  =  (\frac {dN^{2 \to 2}_{ini}}{dyd^2p_{\bot}} (S_{a/A}\ne 1)
             + \frac {dN_{pre}}{dyd^2p_{\bot}}
             + \frac {dN_{the}}{dyd^2p_{\bot}} )
     / \frac {dN^{2 \to 2}_0}{dyd^2p_{\bot}} (S_{a/A}\ne 1)
\end{equation}

To understand the nuclear effect on parton distributions and the roles of the
partonic system, we
need to compare $J/\psi$ production in the central A+B collision with that in
the nucleon-nucleon collision. To this end, a ratio is defined as
\begin{equation}
R  =  (\frac {dN^{2 \to 2}_{ini}}{dyd^2p_{\bot}} (S_{a/A}\ne 1)
             + \frac {dN_{pre}}{dyd^2p_{\bot}}
             + \frac {dN_{the}}{dyd^2p_{\bot}} )
     / \frac {dN^{2 \to 2}_0}{dyd^2p_{\bot}} (S_{a/A}=1)
\end{equation}
which is also written as
\begin{equation}
R=R^{ini}R^{plasma}
\end{equation}

The ratios $R^{ini}$, $S^{plasma}$, $R^{plasma}$ and $R$ versus transverse
momentum and rapidity are depicted as dashed, dotted, dot-dashed and
solid lines, respectively, in Figs. 13 and 14 for the RHIC energy and Figs. 15
and 16 for the LHC energy. In contrast to $R^{ini}<1$, the value of
$R^{plasma}$ is larger than 1 for all transverse momenta in Fig. 13 and
$0<y<1.5$ in Fig. 14 and $0.5<y<1.5$ in Fig. 16. This results in prominent
bulges on the solid lines of $R$ in Figs. 14
and  16. In contrast,
the survival probability $S^{plasma}$
shown by the dotted lines has no such bulge. Therefore, the bulges are present
in Figs. 6 and 10 when the $J/\psi$ yield resulting from
$c\bar c$ pairs produced in the partonic system overwhelms the reduced amount
of initially produced $J/\psi$.
We conclude that in the rapidity region $0<y<1.5$ a bulge observed in
the ratio $R$ is an indicator for the existence of the partonic system.
For $R^{ini}<1$ and $S^{plasma}<1$, $J/\psi$ suppression arises from the
nuclear
parton shadowing found in HIJING and $c\bar c$ dissociation in the partonic
system.

\vspace{0.5cm}
\leftline{\it 6.4. $J/\psi$ number distributions with no suppression}
\vspace{0.5cm}
In Subsection 6.2, $J/\psi$ number distributions have been presented while
the $c\bar c$ reduction due to the nuclear parton shadowing in the initial
collision and $c\bar c$ dissociation
in the partonic system are taken into account. In this subsection, the
suppression including
both the reduction and dissociation is omitted in calculations of $J/\psi$
number distributions by setting to 1 all exponentials in Eqs. (2), (6), (8),
(10) and (11). Figs. 17-20 depict these distributions versus transverse
momentum at $y=0$ and rapidity at $p_T=4$ GeV at both RHIC and LHC energies.
We are now ready
to explain the dip within $y<1$ in Fig. 10. This dip disappears in
Fig. 20 where suppression is not considered. Since $R^{ini}$ shown by the
dashed line in Fig. 16 is
flat with respect to the rapidity $y<2$ and $S^{plasma}$ shown by the
dotted line has a steep rise in $0.5<y<1$, the
dip phenomenon is solely due to the $c \bar c$ dissociation in the partonic
system.
Such a dip phenomenon is not obvious but still can be observed in the
prethermal and
thermal stages when the $J/\psi$ number distributions with suppression are
compared
to those without suppression. The comparison is indicated in Fig. 21 for the
prethermal stage and Fig. 22 for the thermal stage.
The solid and dot-dashed lines for no suppression begin to fall from $y=0.5$
to $y=1$, but change to
rising as shown by the dashed and dotted lines when the $c\bar c$ dissociation
is switched on. This change occurs
because of the steep rise of $S^{plasma}$. A relatively weak
dependence of $S^{plasma}$ on $p_T$ is shown by the dotted line in Fig. 15.
The dip phenomenon thus cannot be observed in the $p_T$
dependence of $J/\psi$ number distributions. Referring back to Eqs. (2), (6),
(8), (10) and (11), exponentials there have sensitive dependence on the
rapidity in
$0<y<1.5$. The dip is more obvious in the color singlet channel as shown by the
lower solid, dashed, dot-dashed and dotted lines in Fig. 12. This is so because
the cross section for
the $1 ^3S_1^{(1)}$-state dissociation has a narrower peak with respect to the
incident gluon energy [12] than the color octet states.

\vspace{0.5cm}
\leftline{\it 6.5. Uncertainties}
\vspace{0.5cm}
Since gluon shadowing in nuclei has not been studied experimentally,
theoretical estimates of the nuclear gluon shadowing factor involve
uncertainties.
The nuclear parton shadowing factor found in HIJING [4] is a result of the
assumption that there is no $Q$-dependence on the shadowing factor and the
shadowing effect for gluons and quarks is the same. Nevertheless, the shadowing
factor has been shown by Eskola {\it et al.} to evolve with momentum $Q$ [39].
The difference between the latter and the former indicates uncertainty.
The ratio $R^{ini}$ defined in Eq. (26) is calculated
with Eskola {\it et al.}'s parametrization [39] and results are depicted in
Fig. 23 showing momentum dependence at $y=0$ and Fig. 24 showing rapidity
dependence at $p_T=4$ GeV. Compared to the dashed lines in Figs. 13-16, the
change of $R^{ini}$ at RHIC energy greater than 1 is
prominent. This
implies that the anti-shadowing effect of Eskola {\it et al.}'s
parametrization is quite
important at RHIC energy. Measurements on $R^{ini}$ in RHIC experiments are
needed to confirm this nuclear enhancement [21].

The ratio $R^{ini}$ is always flat within the rapidity region
-1.5 $< y <$ 1.5 for parametrizations given in HIJING and by Eskola 
{\it et al.},
and the flatness seems to be independent of parametrizations. If the partonic
system does not come into being, the ratio $R$ is flat, too, since
$R = R^{ini}$. If the partonic system dissociates $c\bar c$ pairs, the solid
curve of $R$ undergoes bulging, dipping and then bulging
from $y=-1.5$ to $y=1.5$. Any such twist of $R$ in $-1.5 <y< 1.5$ observed
in experiments is nontrivial, because only a deconfined medium generates it.

Upon inclusion of uncertainties on the formation and evolution of parton
plasma
arising from other factors, for instance, the dependence on the coupling
constant $\alpha_s$ [40] and transverse flow [41], the $J/\psi$
number distribution and the four ratios including survival probability
will change. In the
partonic system considered here gluons dominate the evolution and gluon-$c\bar
c$ interactions break the pairs. In a system where quarks and antiquarks are
abundant,
interactions between quarks (antiquarks) and $c\bar c$ may account for a
suppression of $J/\psi$ [42]. Additional
suppression caused by energy loss of the initial state has not been considered
since there is a controversy on the influence of the energy loss [43, 44].
Some uncertainties are expected to be fixed by upcoming experiments at RHIC.

\vspace{0.5cm}
\leftline{\bf 7. Conclusions}
\vspace{0.5cm}
We have studied $J/\psi$ production through both color-singlet and color-octet
$c\bar c$ channels with various stages of central Au-Au collisions at both RHIC
and LHC energies. In addition to
the scattering processes $ab \to c\bar {c}[ ^{2S+1}L_J] {\rm x}$, contributions
of the reactions $ab \to c\bar {c}[ ^{2S+1}L_J]$ are also calculated in the
prethermal stage and thermal stage.
The effect of the medium on an expanding $c\bar c$ involves a gluon
interacting with the $c\bar c$ to prevent it from a transition into a color
singlet. Cross sections for
$g+c\bar {c}
\to (c\bar c)_8$ are calculated with internal wave functions of $(c\bar c)_1$
in an attractive potential and $(c\bar c)_8$ in a repulsive
potential.
Furthermore, nucleon-$c\bar {c}$ cross sections for color singlet, $S$- and
$P$- wave color octets as a function of $\sqrt s$ or $c\bar
c$ radius are evaluated by assuming that the nucleon dominantly contains
gluons.
Momentum and rapidity dependence of $J/\psi$ number distribution with various
contributions are calculated for central Au-Au collisions at both RHIC and
LHC energies. Color octet contributions are one order of magnitude larger than
the color singlet contributions. Yields of $c\bar c$ are large in
the prethermal
stage at RHIC energy and through the  $2 \to 1$ collisions $ab \to
c\bar {c}[ ^{2S+1}L_J]$ at LHC energy. Since the partonic system
offers fairly large amounts of $c\bar c$, a bulge in $0<y<1.5$ at RHIC
energy and $0.5<y<1.5$ at LHC energy can be observed in the rapidity
dependence of the $J/\psi$ number distribution and the ratio $R$ of $J/\psi$
number distributions for Au-Au collisions to nucleon-nucleon collisions. Such
a bulge is
a signature for the existence of a deconfined partonic medium. We suggest
that RHIC and LHC experiments measure $J/\psi$ number
distributions and the ratio $R$ in the rapidity region $0<y<3$ to observe
a bulge. While the yield of
$c\bar c$ from the medium is larger than the reduced amount of initial
production in the medium,
the ratio $R^{plasma}$ is larger than 1. The competition between
production and suppression determines the values of $R^{plasma}$, which
relies on the evolution of parton number density and temperature of the
partonic system [12]. A dip in the rapidity dependence of the $J/\psi$ number
distributions at LHC energy may exist and this amounts to a suppression
effect of $c\bar c$ in the partonic system. So far, we have obtained results
and conclusions
for positive rapidity. It is stressed that the same contents for
negative rapidity can be obtained from the positive region by symmetry.

\vspace{0.5cm}
\leftline{\bf Acknowledgements}
\vspace{0.5cm}
I thank the [Department of Energy's] Institute for Nuclear Theory
at the University of Washington
for its hospitality and the Department of Energy for partial support during the
completion of this work. I thank the Nuclear Theory Group at LBNL
Berkeley for their hospitality during my visit. I also thank X.-N. Wang,
C.-Y. Wong and
M. Asakawa for discussions, K. J. Eskola for offering Fortran codes of
nuclear parton shadowing factors, H. J. Weber for careful reading through the
manuscript. This work was also supported
in part by the project KJ951-A1-410 of the Chinese Academy of Sciences and the
Education Bureau of Chinese Academy of Sciences.

\newpage
\centerline{\bf References}
\vskip 14pt
\leftline{[1]D. Kharzeev and H. Satz, Phys. Lett. B334(1994)155.}
\leftline{[2]R. C. Hwa and K. Kajantie, Phys. Rev. Lett. 56(1986)696;}
\leftline{~~~J. P. Blaizot and A. H. Mueller, Nucl. Phys. B289(1987)847.}
\leftline{[3]K. Kajantie, P. V. Landshoff, and J. Lindfors, Phys. Rev. Lett.
59}
\leftline{~~~(1987)2517;}
\leftline{~~~K. J. Eskola, K. Kajantie, and J. Lindfors, Nucl. Phys.
             B323(1989)37;}
\leftline{~~~Phys. Lett. B214(1991)613.}
\leftline{[4]X.-N. Wang and M. Gyulassy, Phys. Rev. D44(1991)3501; Comput.}
\leftline{~~~Phys. Commun. 83(1994)307.}
\leftline{~~~X.-N. Wang, Phys. Rep. 280(1997)287.}
\leftline{[5]K. Geiger and B. M$\rm {\ddot u}$ller, Nucl. Phys.
B369(1992)600;}
\leftline{~~~K. Geiger, Phys. Rev. D47(1993)133.}
\leftline{[6]H. J. Moehring and J. Ranft, Z. Phys. C52(1991)643;}
\leftline{~~~P. Aurenche et al., Phys. Rev. D45(1992)92;}
\leftline{~~~P. Aurenche et al., Comput. Phys. Commun. 83(1994)107.}
\leftline{[7]E. Shuryak, Phys. Rev. Lett. 68(1992)3270;}
\leftline{~~~L. Xiong and E.Shuryak,
Phys. Rev. C49(1994)2203.}
\leftline{[8]K. Geiger and J. I. Kapusta, Phys. Rev. D47(1993)4905.}
\leftline{[9]T. S. Bir$\rm {\acute o}$, E. van Doorn, B.
M$\rm {\ddot u}$ller, M. H. Thoma, and X.-N. Wang,}
\leftline{~~~Phys. Rev. C48(1993)1275.}
\leftline{[10]J. Alam, S. Raha and B. Sinha, Phys. Rev. Lett. 73(1994)1895.}
\leftline{[11]H. Heiselberg and X.-N. Wang, Phys. Rev. C53(1996)1892.}
\leftline{[12]X.-M. Xu, D. Kharzeev, H. Satz and X.-N. Wang, Phys. Rev.
C53}
\leftline{~~~~(1996)3051.}
\leftline{[13]T. Matsui and H. Satz, Phys. Lett. B178(1986)416.}
\leftline{[14]D. Kharzeev and H. Satz, Phys. Lett. B366(1996)316.}
\leftline{[15]D. M. Alde et al., Phys. Rev. Lett. 66(1991)133;}
\leftline{~~~~D. M. Alde et al., Phys. Rev. Lett. 66(1991)2285;}
\leftline{~~~~L. Antoniazzi et al., Phys. Rev. Lett. 70(1993)383;}
\leftline{~~~~M. H. Schub et al., Phys. Rev. D52(1995)1307;}
\leftline{~~~~T. Alexopoulos et al., Phys. Rev. D55(1997)3927}
\leftline{[16]F. Abe et al., CDF Collaboration, Phys. Rev. Lett.
79(1997)572,578}
\leftline{[17]W. E. Caswell and G. P. Lepage, Phys. Lett. B167(1986)437;}
\leftline{~~~~G. P. Lepage, L. Magnea, C. Nakhleh, U. Magnea and K.
Hornbostel,}
\leftline{~~~~Phys. Rev. D46(1992)4052;}
\leftline{~~~~G. T. Bodwin, E. Braaten and G. P. Lepage, Phys. Rev.
D51(1995)1125.}
\leftline{[18]P. L${\rm {\acute {e}}}$vai, B. M${\rm {\ddot {u}}}$ller and
X.-N. Wang, Phys. Rev. C51(1995)3326.}
\leftline{[19]Z. Lin and M. Gyulassy, Phys. Rev. C51(1995)2177.}
\leftline{[20]K. J. Eskola and X.-N. Wang, Phys. Rev. D49(1994)1284.}
\leftline{[21]Y. Akiba, in Proc. of
Charmonium Production in Relativistic Nuclear}
\leftline{~~~~Collisions, INT,
Seattle, 1998, eds. B. Jacak and X.-N. Wang (World}
\leftline{~~~~Scientific, Singapore,1998);}
\leftline{~~~~M. Rosati, in Proc. of
Charmonium Production in Relativistic Nuclear}
\leftline{~~~~Collisions, INT,
Seattle, 1998, eds. B. Jacak and X.-N. Wang (World}
\leftline{~~~~Scientific, Singapore,1998)}
\leftline{[22]S. Gupta and H. Satz, Z. Phys. C55(1992)391.}
\leftline{~~~~R. C. Hwa and L. Le$\rm \acute s$niak, Phys. Lett. B295(1992)11.}
\leftline{~~~~R. Vogt, S. J. Brodsky and P. Hoyer, Nucl Phys. B360(1991)67;}
\leftline{~~~~K. Boreskov, A. Capella, A. Kaidalov and J. Tran Thanh Van,
Phys.}
\leftline{~~~~Rev. D47(1993)919.}
\leftline{~~~~M. A. Braun, C. Pajares, C. A. Salgado, N. Armesto and A.
Capella,}
\leftline{~~~~Nucl. Phys. B509(1998)357.}
\leftline{[23]P. Cho and A. K. Leibovich, Phys. Rev. D53(1996)150,6203.}
\leftline{[24]K. Sridhar, A. D. Martin and W. J. Stirling, Phys. Lett.
B438(1998)211.}
\leftline{[25]R. Baier and R. R${\rm {\ddot {u}}}$ckl, Z. Phys. C19(1983)251;}
\leftline{~~~~R. Gastmans, W. Troost and T. T. Wu, Nucl. Phys. B291(1987)731.}
\leftline{[26]M. E. Peskin, Nucl. Phys. B156(1979)365;}
\leftline{~~~~G. Bhanot and M. E. Peskin, Nucl. Phys. B156(1979)391.}
\leftline{[27]B. Z. Kopeliovich and B. G. Zakharov, Phys. Rev. D44(1991)3466;}
\leftline{~~~~L. Frankfurt, G. A. Miller and M. Strikman, Phys. Lett.
B304(1993)1;}
\leftline{~~~~L. Gerland, L. Frankfurt, M. Strikman, H. St$\rm \ddot o$cker and
W. Greiner,}
\leftline{~~~~Phys. Rev. Lett. 81(1998)762.}
\leftline{~~~~P. Jain, B. Pire and J. P. Ralston, Phys. Rep. 271(1996)67.}
\leftline{[28]M. Gl$\rm \ddot {u}$ck, E. Reya and A. Vogt, Z. Phys.
C67(1995)433.}
\leftline{[29]R. L. Anderson, SLAC-Pub 1741(1976).}
\leftline{[30]J. H$\rm \ddot u$fner and B. Z. Kopeliovich, Phys.
Lett. B426(1998)154.}
\leftline{[31]C. Gerschel and J. H$\rm \ddot u$fner, Z. Phys. C56(1992)171.}
\leftline{[32]J. Dolej$\rm \breve s$i and J. H$\rm \ddot u$fner, Z. Phys.
C54(1992)489.}
\leftline{~~~~C. W. Wong, Phys. Rev. D54(1996)R4199.}
\leftline{[33]C.-Y. Wong and C. W. Wong, Phys. Rev. D57(1998)1838.}
\leftline{[34]W. Cassing and E. L. Bratkovskaya, Nucl. Phys. A623(1997)570.}
\leftline{[35]M.
Beneke and M. Kr$\rm \ddot {a}$mer, Phys. Rev. D55(1997)R5269.}
\leftline{[36]J. Amundson, S. Fleming and I. Maksymyk, Phys. Rev.
D56(1997)5844;}
\leftline{~~~~T. Mehen, Phys. Rev. D55(1997)4338.}
\leftline{[37]R. Gavai et al., Int. J. Mod. Phys. A10(1995)3043.}
\leftline{[38]A. H. Mueller and J. Qiu, Nucl. Phys. B268(1986)427;}
\leftline{~~~~K. J. Eskola, J. Qiu and X.-N. Wang, Phys. Rev. Lett.
72(1994)36;}
\leftline{~~~~M. Arneodo, Phys. Rep. 240(1994)301.}
\leftline{[39]K. J. Eskola, V. J. Kolhinen and C. A. Salgado, JYFL-8/98,}
\leftline{~~~~US-FT/14-98, hep-ph/9807297.}
\leftline{~~~~K. J. Eskola, V. J. Kolhinen and P. V. Ruuskanen,
CERN-TH/97-345,}
\leftline{~~~~JYFL-2/98, hep-ph/9802350.}
\leftline{[40]S. M. H. Wong, Phys. Rev. C56(1997)1075.}
\leftline{[41]D. K. Srivastava, M. G. Mustafa and B. M$\rm \ddot u$ller, Phys.
Rev. C56(1997)1064.}
\leftline{[42]R. Wittmann and U. Heinz, Z. Phys. C59(1993)77.}
\leftline{[43]S. Gavin and J. Milana, Phys. Rev. Lett. 68(1992)1834;}
\leftline{~~~~E. Quack and T. Kodama, Phys. Lett. B302(1993)495;}
\leftline{~~~~R. C. Hwa, J. Pi$\rm \breve s$$\rm \acute u$t and N. Pi$\rm
\breve s$$\rm \acute u$tov$\rm \acute a$, Phys. Rev. C56(1997)432.}
\leftline{[44]S. J. Brodsky and P. Hoyer, Phys. Lett. B298(1993)165.}

\newpage 
\begin{figure}[t] 
  \begin{center} 
     \includegraphics[width=0.65\textwidth,angle=0]{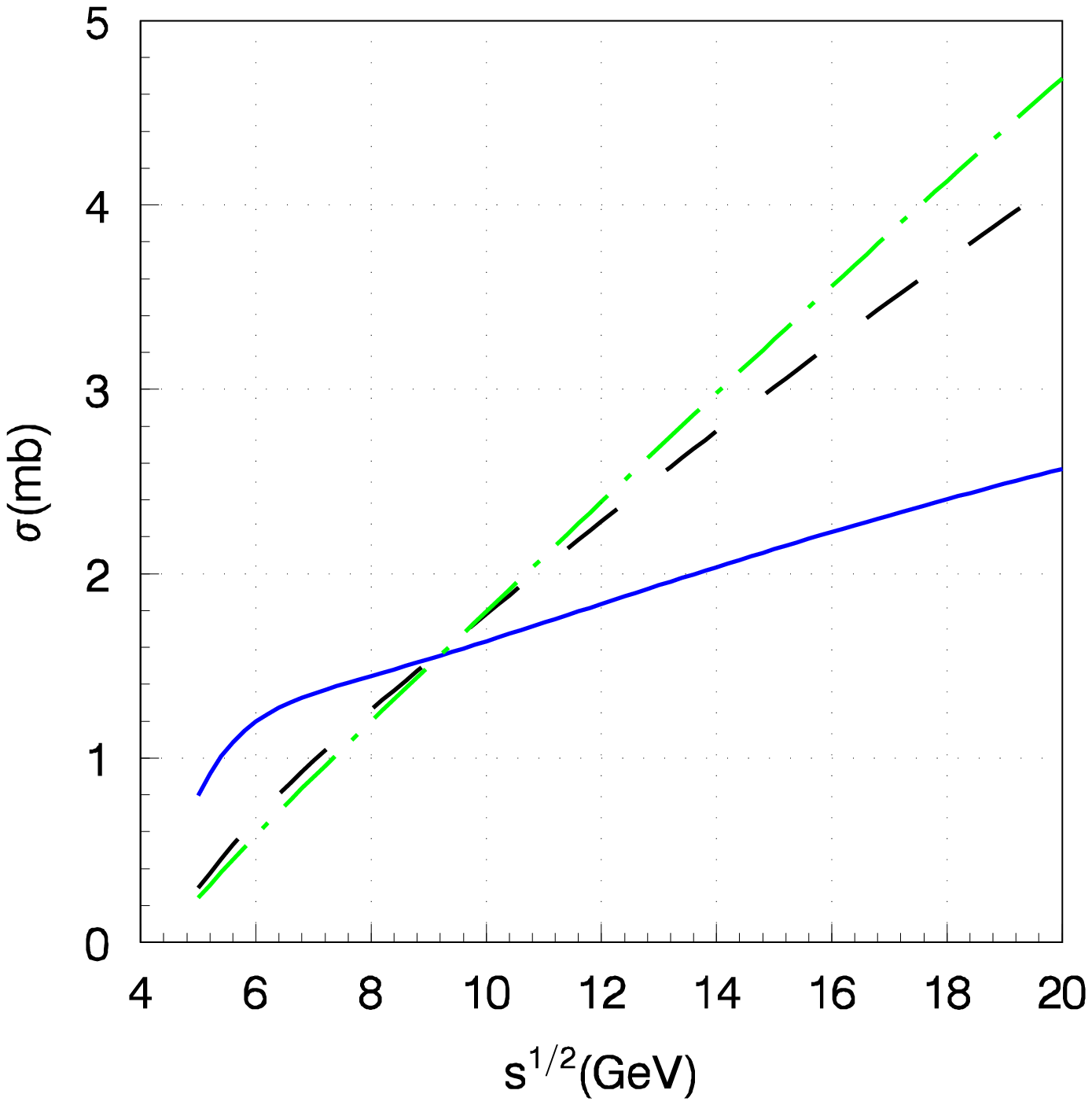}
  \end{center} 
\caption{Solid, dashed and dot-dashed lines are nucleon-$c\bar {c}[1
^3S_1^{(1)}]$ cross
sections for $f_{g/N}(x, Q^2)$ evaluated at $Q^2=\epsilon_0^2, 2\epsilon_0^2,
(Q^0)^2$, respectively. The $(c\bar {c})_1$ has the same size as $J/\psi$.}  
\label{fig1} 
  \begin{center} 
     \includegraphics[width=0.65\textwidth,angle=0]{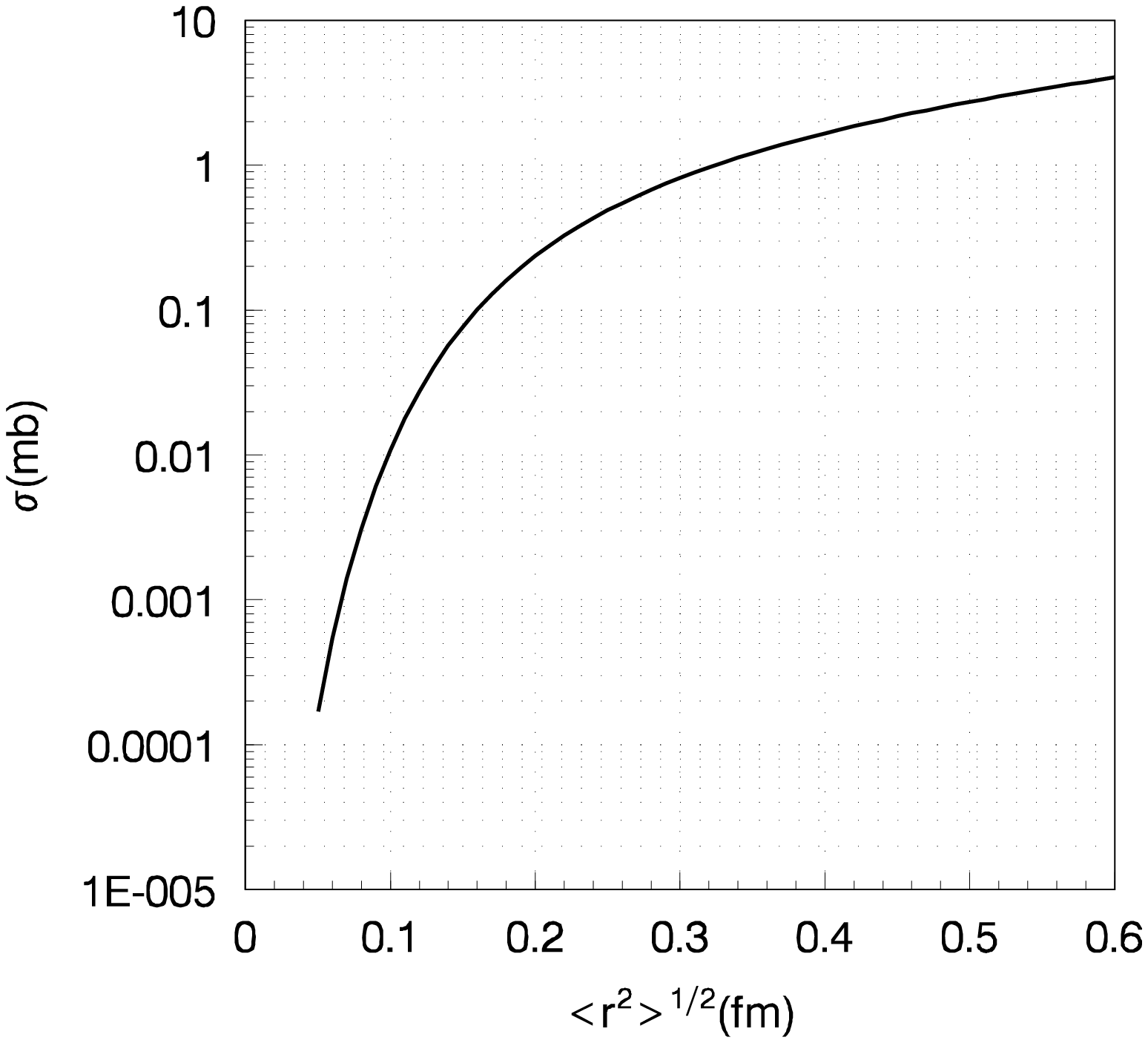}
  \end{center} 
\caption{Cross section for nucleon-$c\bar {c}[1 ^3S_1^{(1)}]$ at $\sqrt
{s}=6$ GeV as a function of the $c\bar c$
radius is calculated with $f_{g/N}(x, Q^2)$ evaluated at $Q^2=\epsilon_0^2$.}  
\label{fig2} 
\end{figure} 

\newpage 
\begin{figure}[t] 
  \begin{center} 
     \includegraphics[width=0.65\textwidth,angle=0]{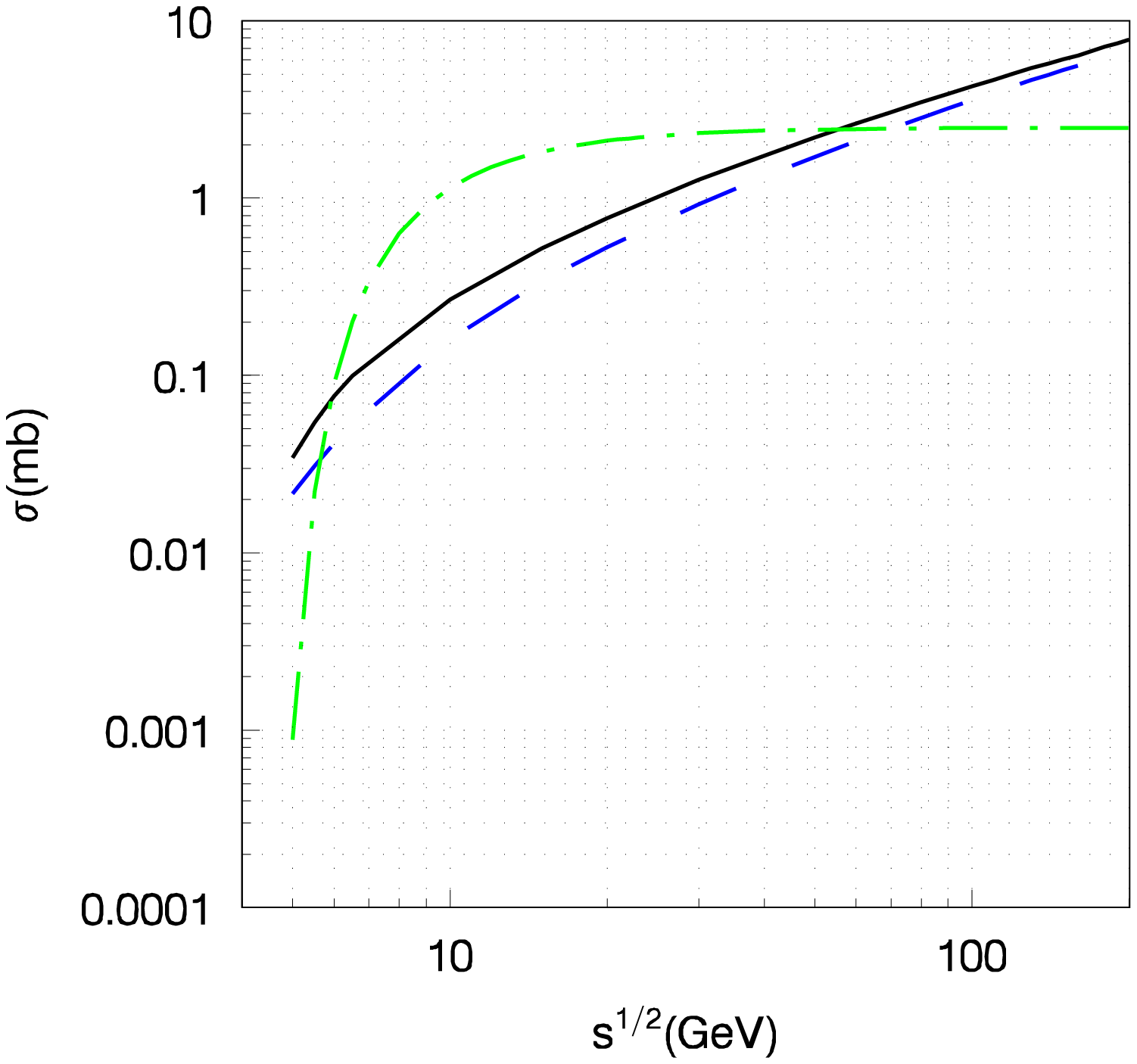}
  \end{center} 
\caption{The solid and dashed lines are cross sections for nucleon-$c\bar
c[S^{(8)}]$ and nucleon-$c\bar c[P^{(8)}]$ collisions as a function of $\sqrt
s$, respectively. The dot-dashed line is the nucleon-$J/\psi$ cross section 
calculated with Eq. (24) in Ref. [1]. The corresponding $(c\bar c)_8$ and 
$J/\psi$ have the same radius.}  
\label{fig3} 
  \begin{center} 
     \includegraphics[width=0.65\textwidth,angle=0]{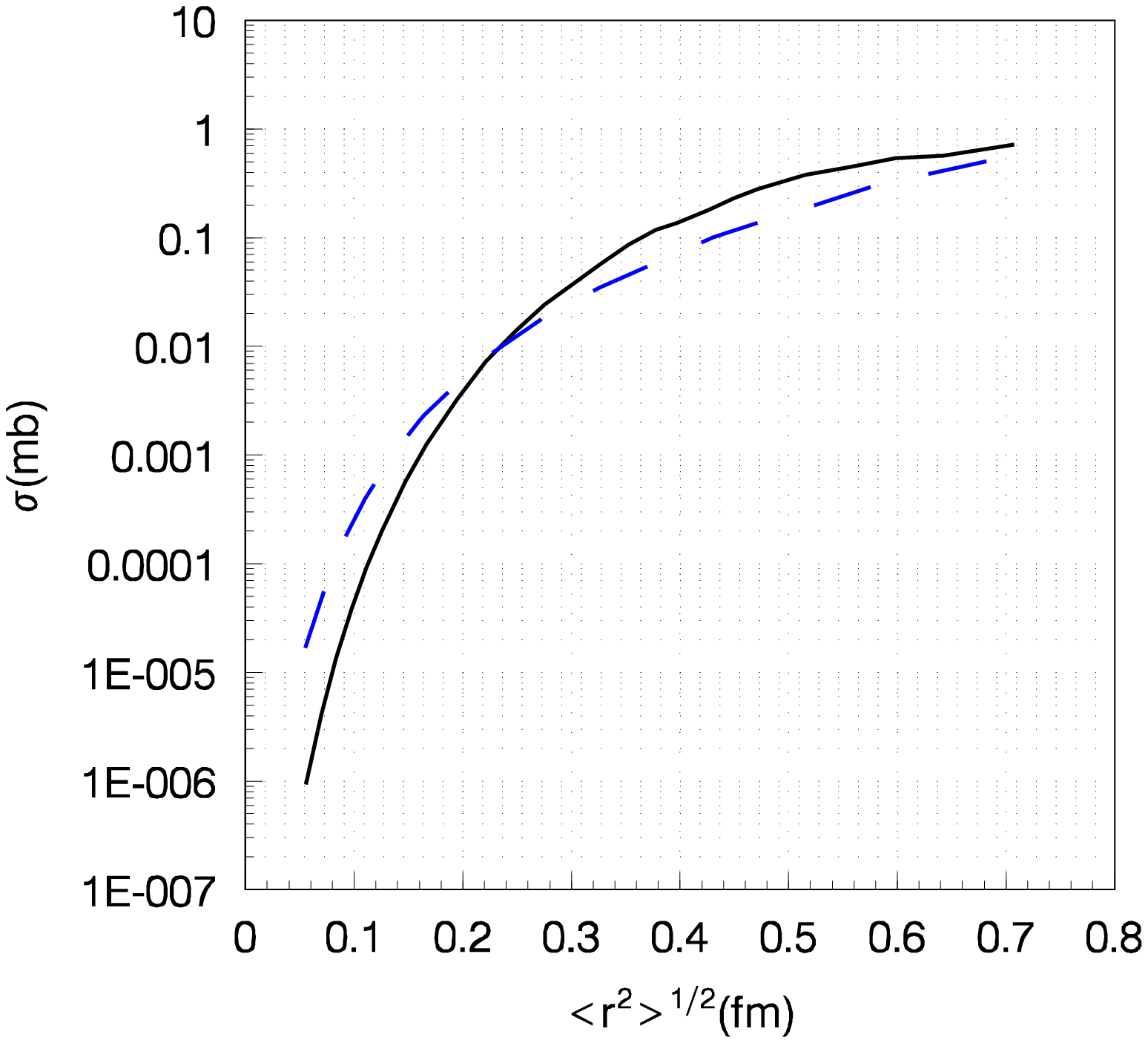}
  \end{center} 
\caption{The solid and dashed lines show radius dependence of cross sections 
for nucleon-$(c\bar c)_8[S^{(8)}]$ and nucleon-$(c\bar c)_8[P^{(8)}]$ 
collisions, respectively.}  
\label{fig4} 
\end{figure} 

\newpage 
\begin{figure}[t] 
  \begin{center} 
     \includegraphics[width=0.9\textwidth,angle=0]{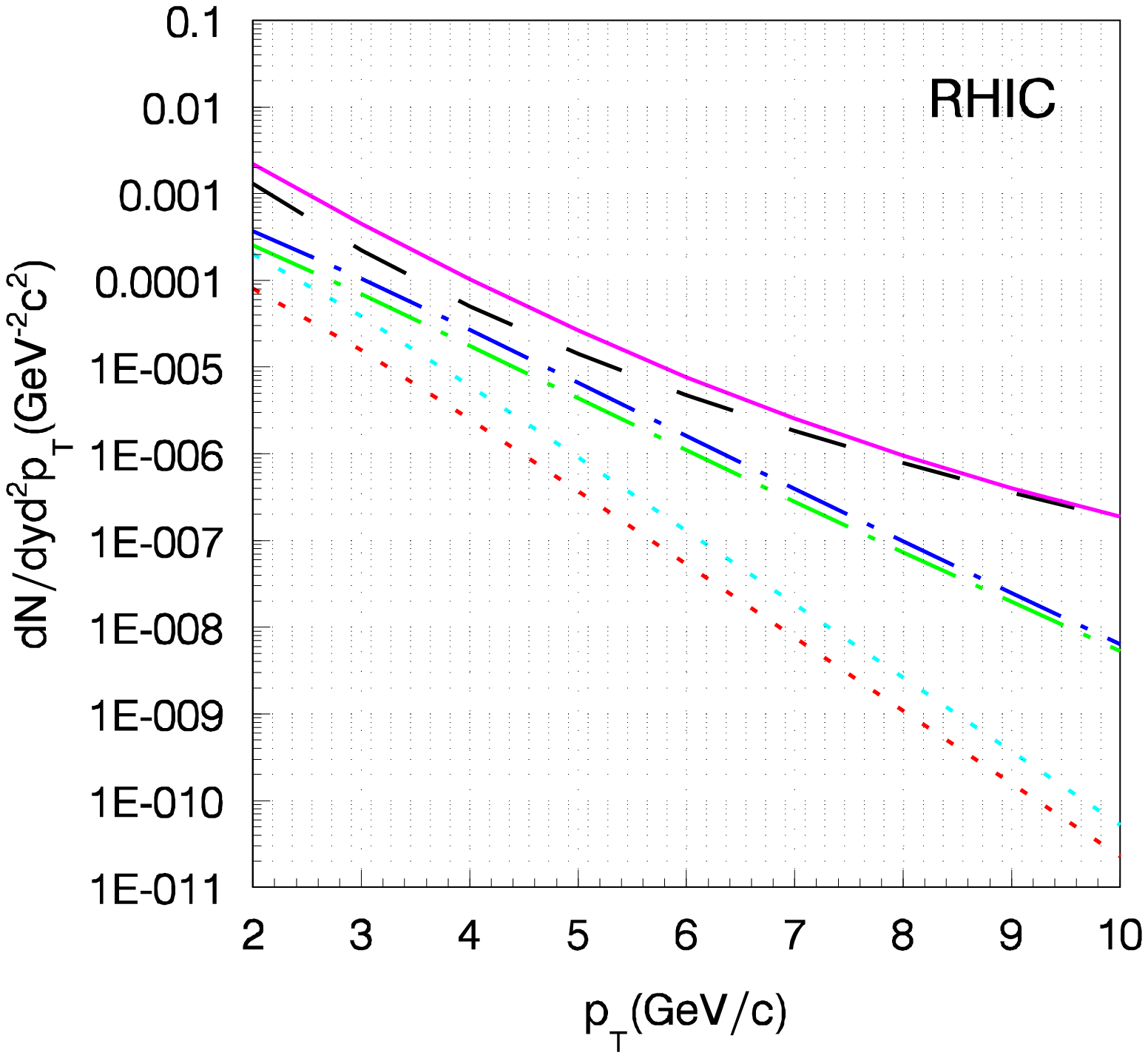}
  \end{center} 
\caption{$J/\psi$ number distributions versus transverse momentum at $y=0$ and
RHIC energy with suppression. The dashed curve corresponds to $c\bar c$ 
production in the initial collision. The upper and lower
dot-dashed (dotted) curves correspond to $c\bar c$ produced through $2 \to 1$
and $2 \to 2$ reactions in the prethermal (thermal) stage, respectively.
The solid curve is the sum of all contributions.}  
\label{fig5} 
\end{figure} 

\newpage 
\begin{figure}[t] 
  \begin{center} 
     \includegraphics[width=0.9\textwidth,angle=0]{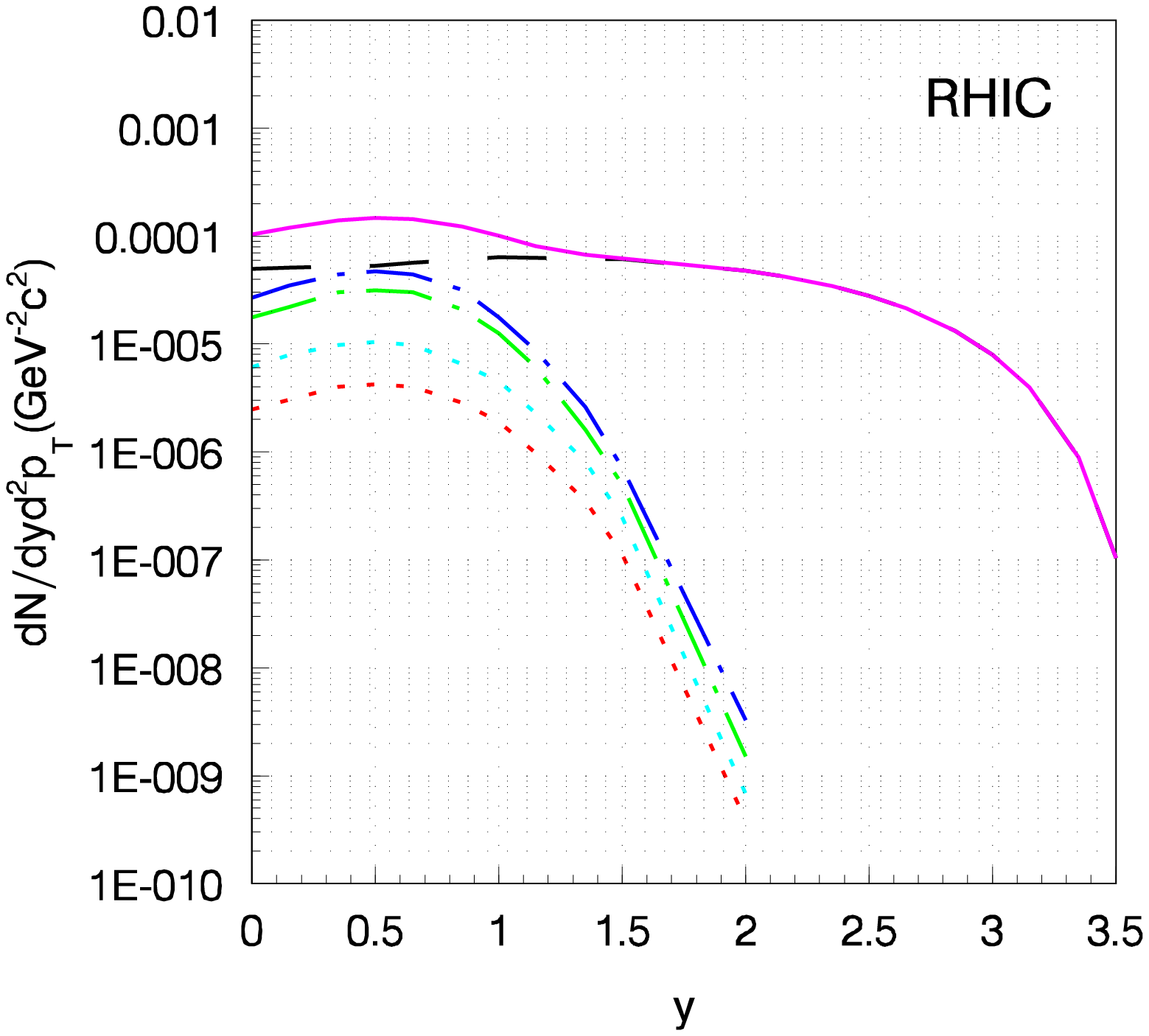}
  \end{center} 
\caption{The same as Fig. 5, except for rapidity distribution at
$p_T=4$ GeV.}  
\label{fig6} 
\end{figure} 

\newpage 
\begin{figure}[t] 
  \begin{center} 
     \includegraphics[width=0.9\textwidth,angle=0]{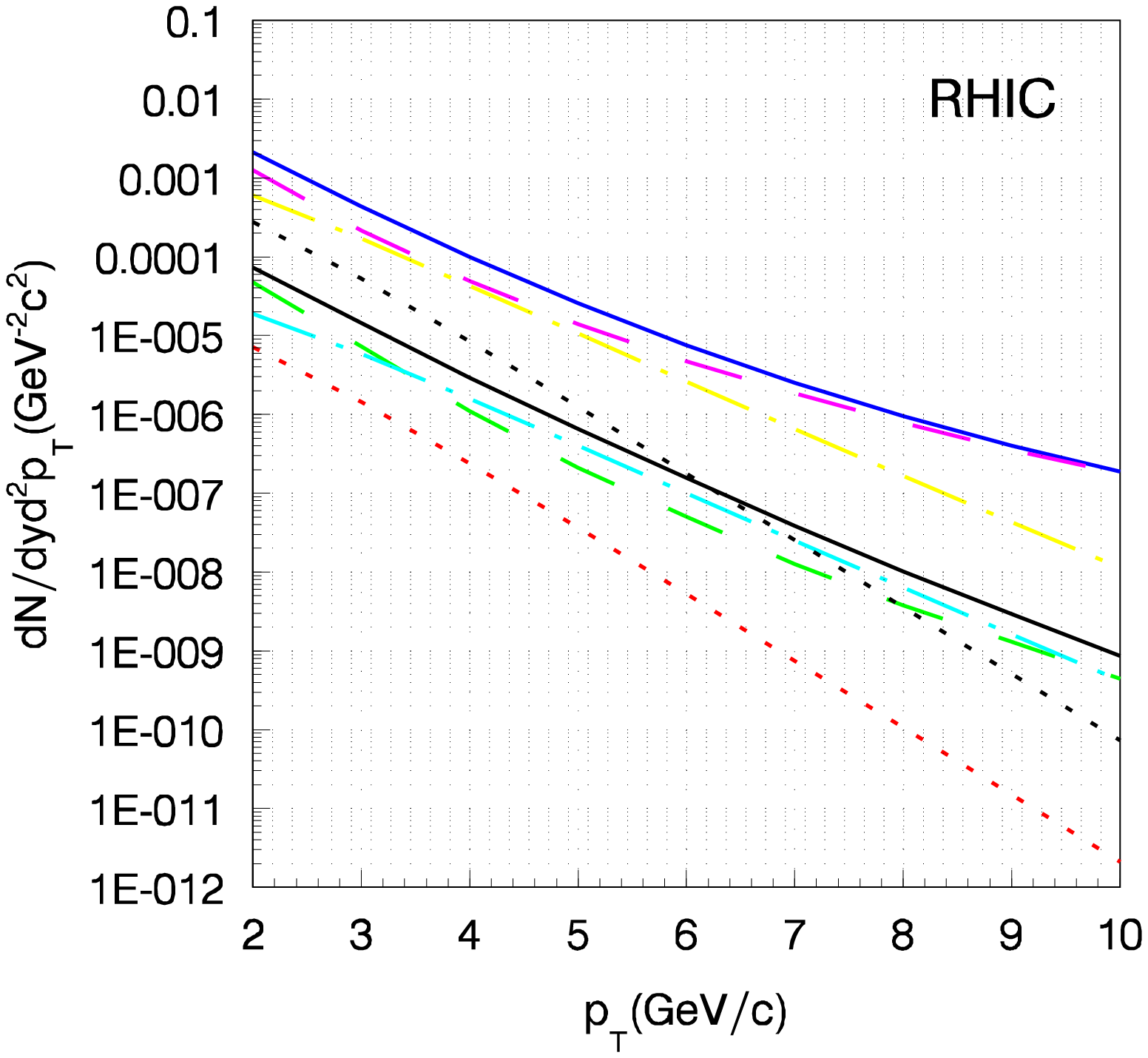}
  \end{center} 
\caption{$J/\psi$ number distributions
versus transverse momentum at $y=0$ and RHIC
energy with suppression. The upper and lower dashed (dot-dashed, dotted and
solid) lines correspond to $c\bar c$ in color octet and color singlet,
respectively, produced in the initial collision(prethermal stage, thermal stage
and the all three stages).}  
\label{fig7} 
\end{figure} 

\newpage 
\begin{figure}[t] 
  \begin{center} 
     \includegraphics[width=0.9\textwidth,angle=0]{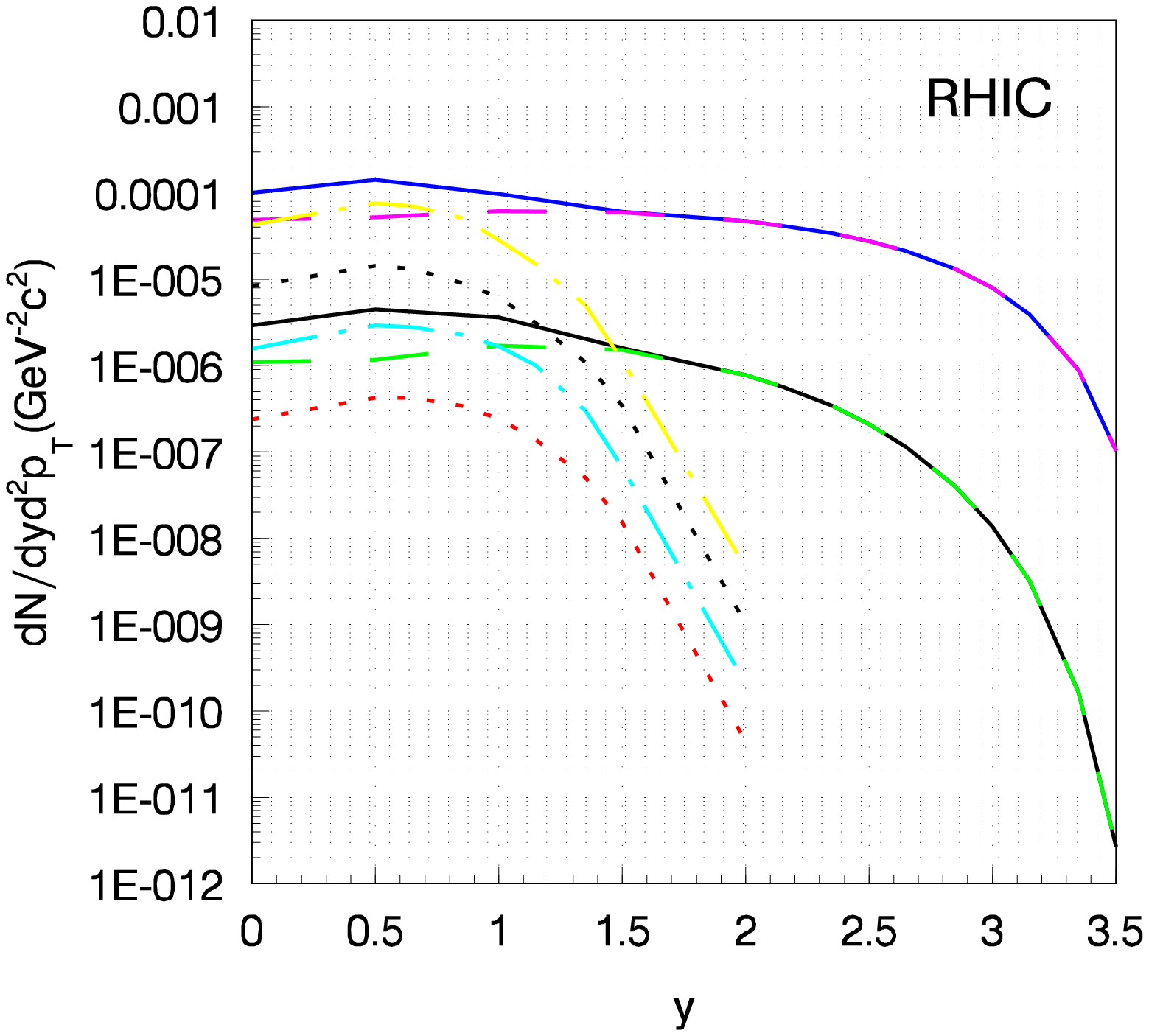}
  \end{center} 
\caption{The same as Fig. 7, except for rapidity distribution at
$p_T=4$ GeV.}  
\label{fig8} 
\end{figure} 

\newpage 
\begin{figure}[t] 
  \begin{center} 
     \includegraphics[width=0.9\textwidth,angle=0]{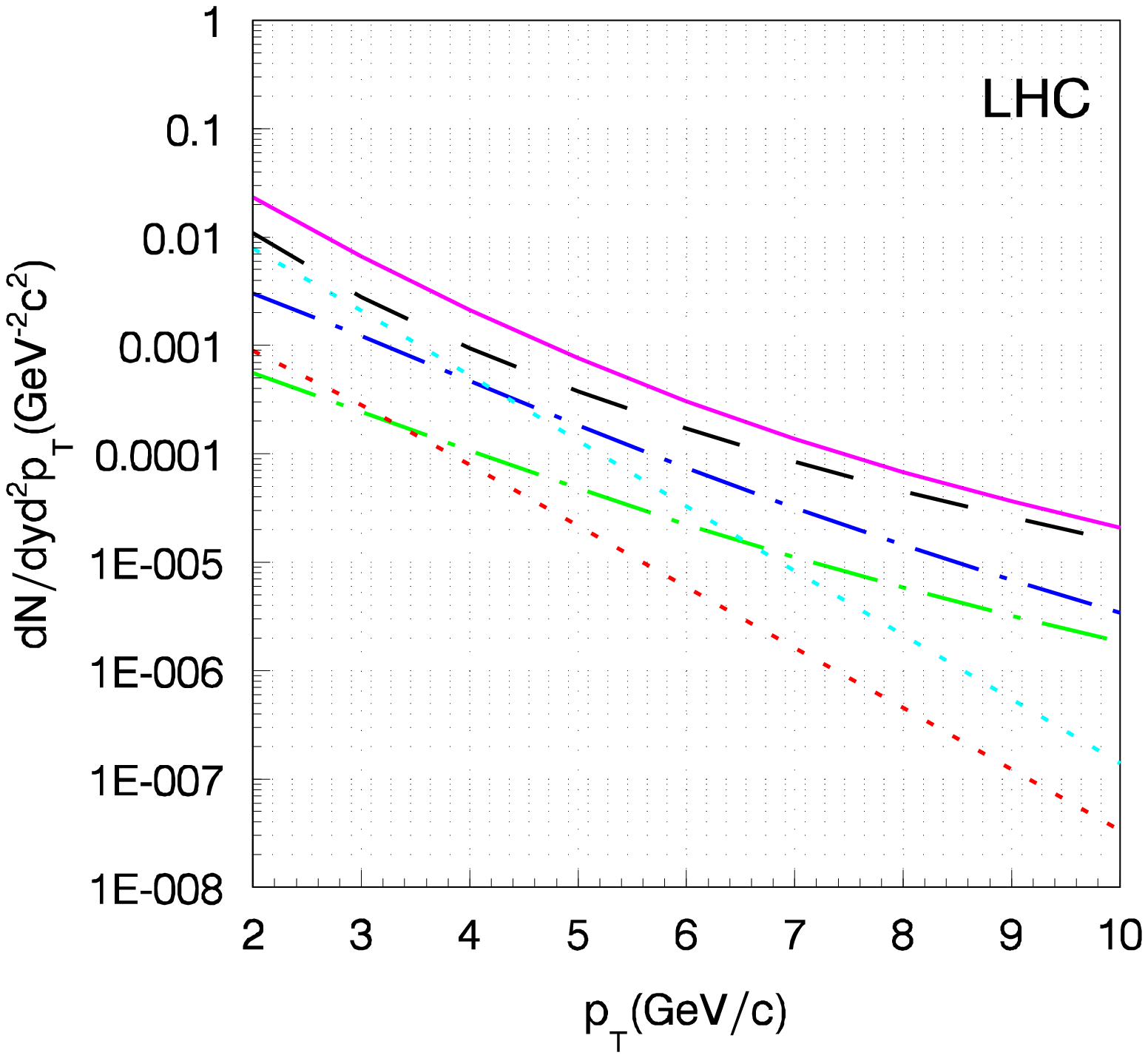}
  \end{center} 
\caption{$J/\psi$ number distributions versus transverse momentum at $y=0$ and
LHC energy with suppression. The dashed curve corresponds to $c\bar c$ 
productions in the initial collision. The upper and lower
dot-dashed (dotted) curves correspond to $c\bar c$ produced through $2 \to 1$
and $2 \to 2$ reactions in the prethermal (thermal) stage, respectively.
The solid curve is the sum of all contributions.}  
\label{fig9} 
\end{figure} 

\newpage 
\begin{figure}[t] 
  \begin{center} 
     \includegraphics[width=0.9\textwidth,angle=0]{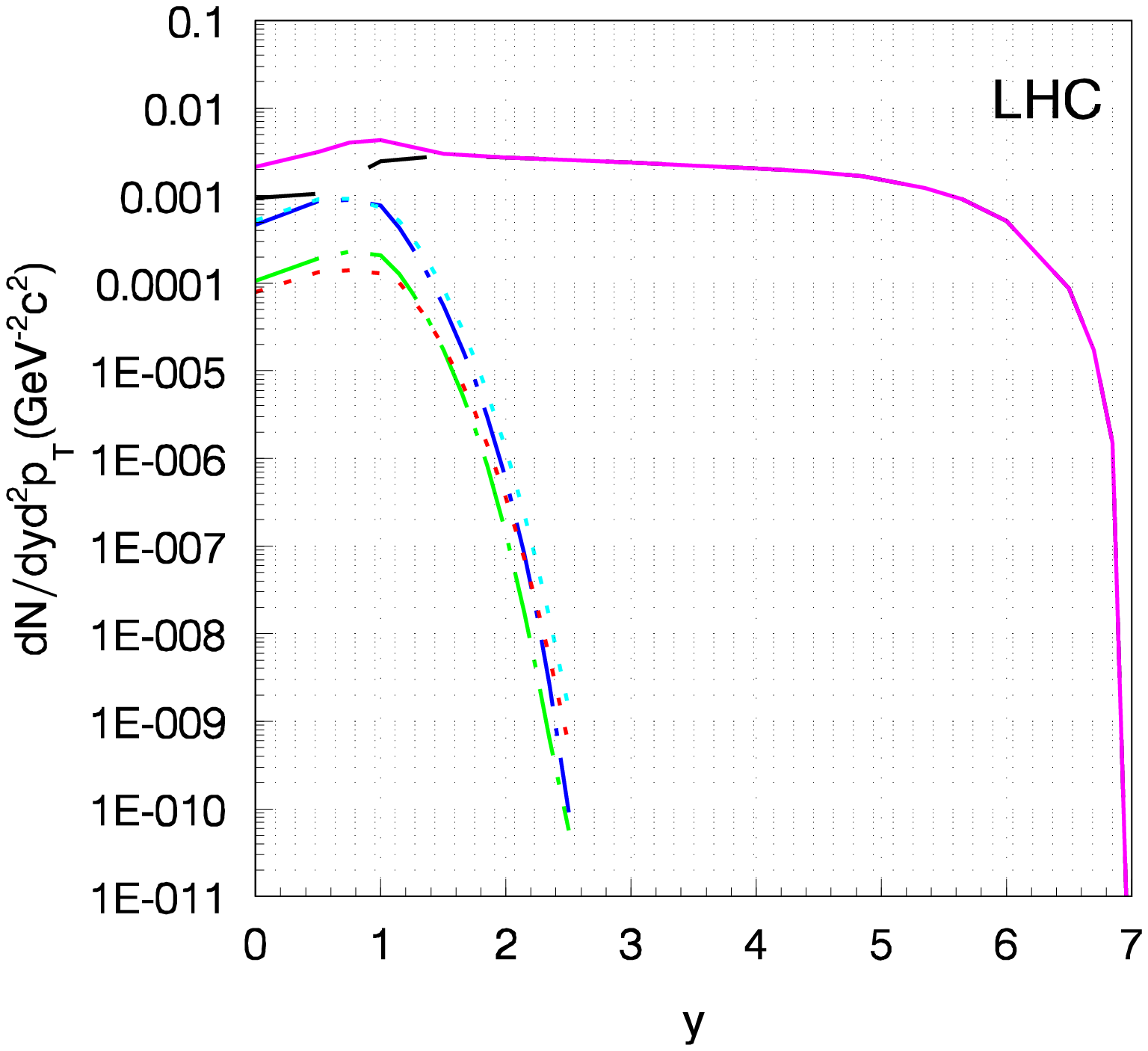}
  \end{center} 
\caption{ The same as Fig. 9, except for rapidity distribution at
$p_T=4$ GeV.}  
\label{fig10} 
\end{figure} 

\newpage 
\begin{figure}[t] 
  \begin{center} 
     \includegraphics[width=0.9\textwidth,angle=0]{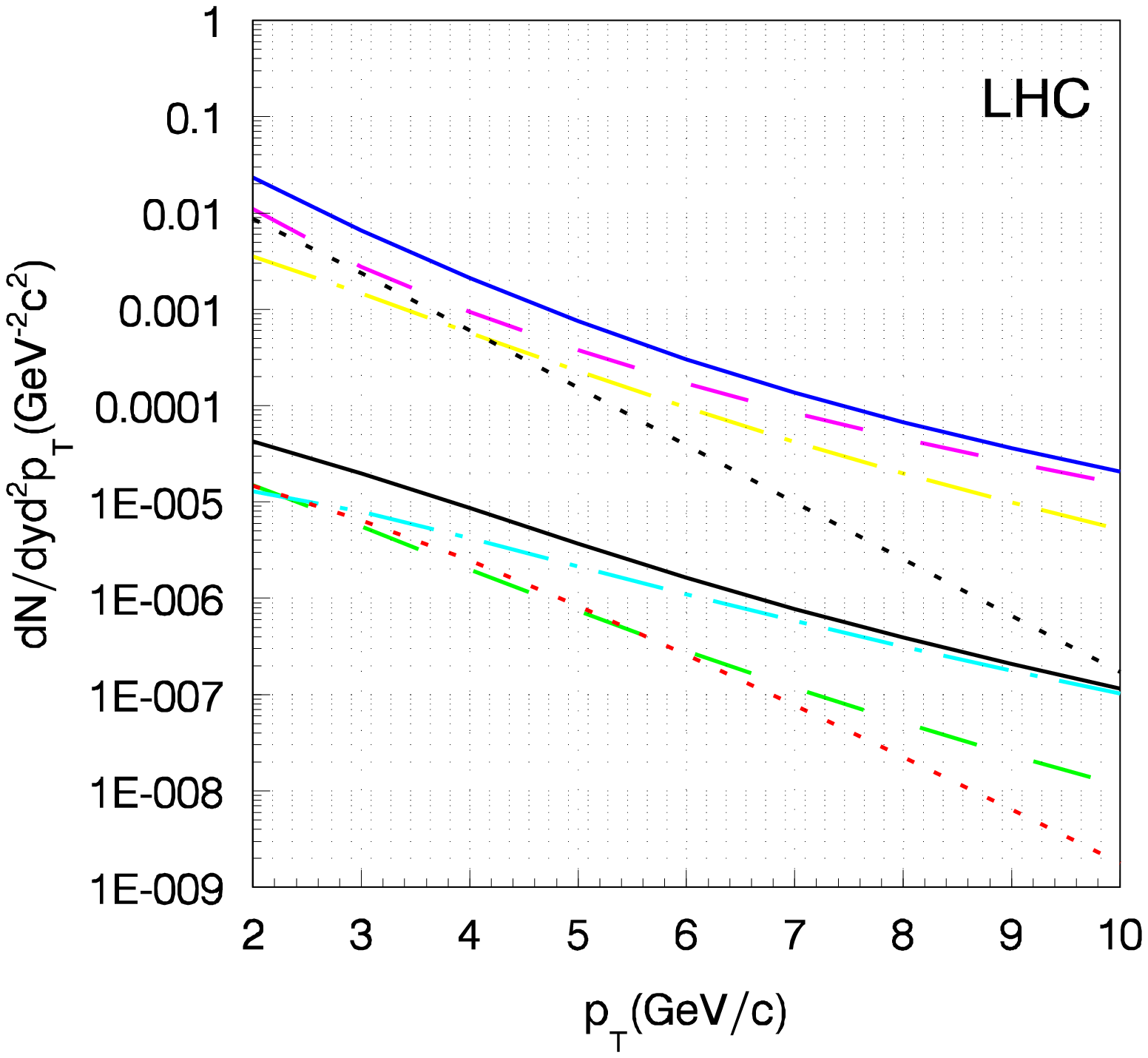}
  \end{center} 
\caption{$J/\psi$ number distributions
versus transverse momentum at $y=0$ and LHC
energy with suppression. The upper and lower dashed (dot-dashed, dotted and
solid) lines correspond to $c\bar c$ in color octet and color singlet,
respectively, produced in the initial collision (prethermal stage, thermal
stage and all three stages).}  
\label{fig11} 
\end{figure} 

\newpage 
\begin{figure}[t] 
  \begin{center} 
     \includegraphics[width=0.9\textwidth,angle=0]{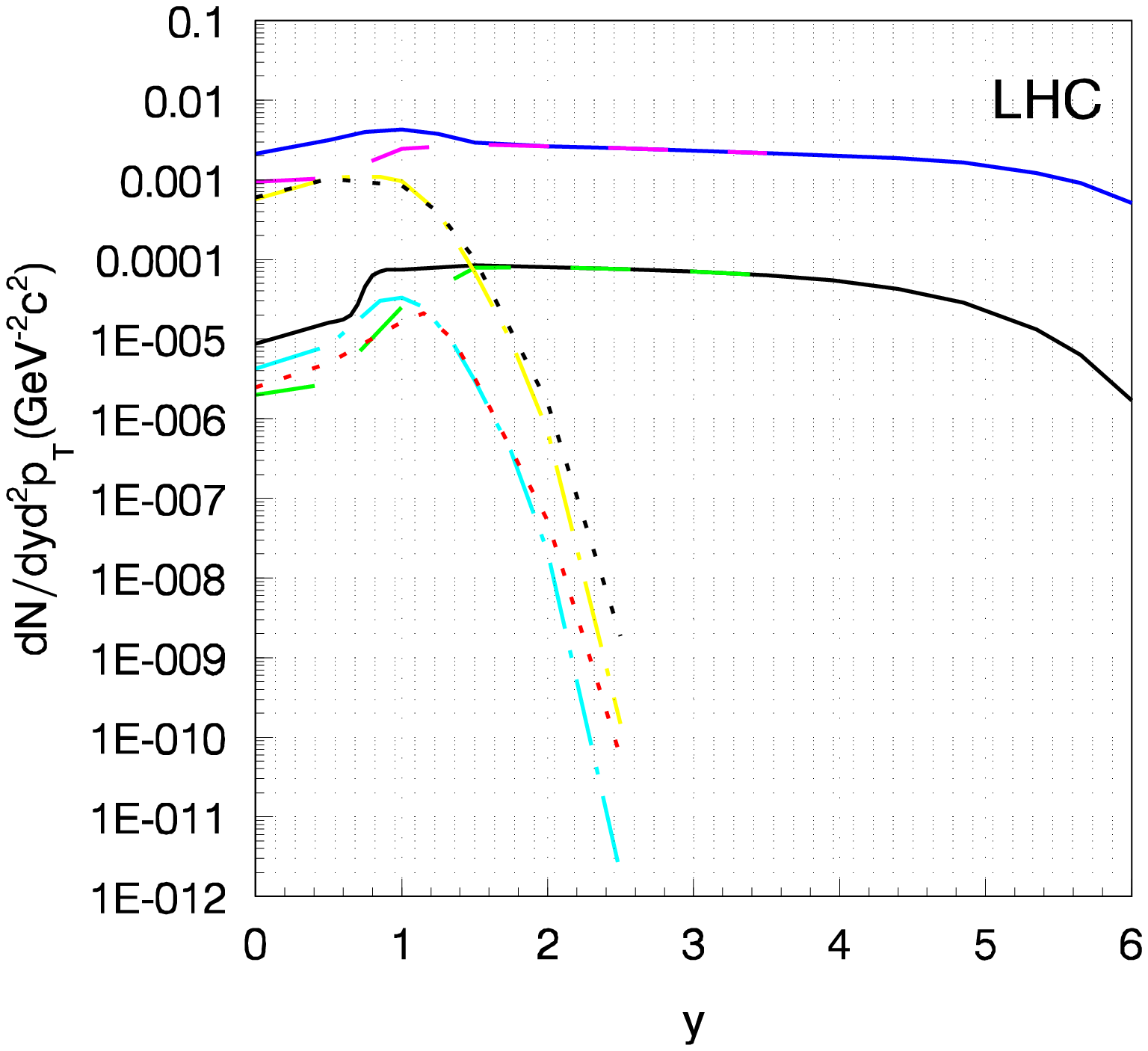}
  \end{center} 
\caption{The same as Fig. 11, except for rapidity distribution at
$p_T=4$ GeV.}  
\label{fig12} 
\end{figure} 

\newpage 
\begin{figure}[t] 
  \begin{center} 
     \includegraphics[width=0.7\textwidth,angle=0]{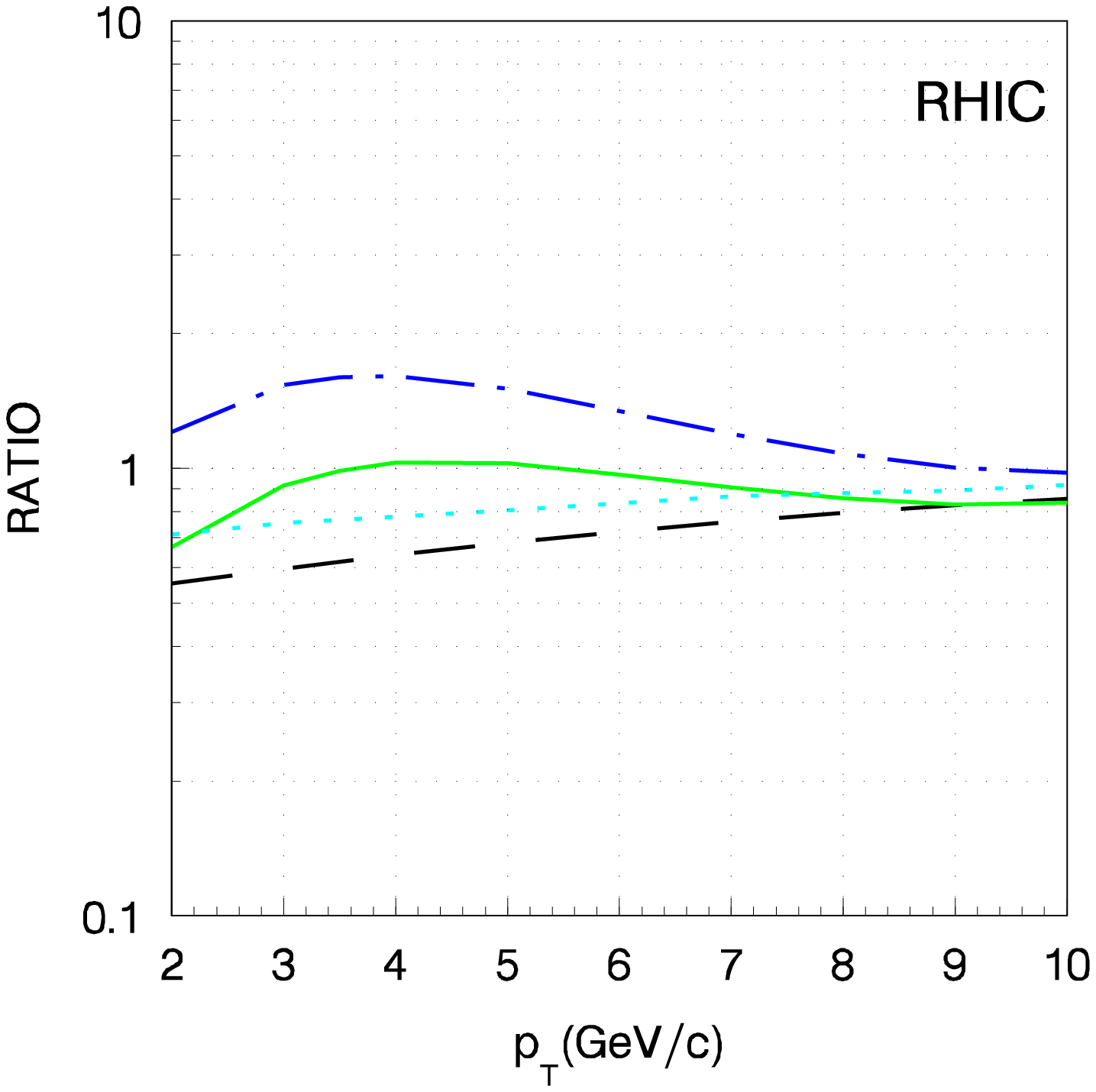}
  \end{center} 
\caption{Ratios versus transverse momentum at $y=0$ and RHIC
energy. The solid, dashed, dot-dashed and dotted lines are
$R$, $R^{ini}$, $R^{plasma}$ and $S^{plasma}$, respectively.}  
\label{fig13} 
  \begin{center} 
     \includegraphics[width=0.7\textwidth,angle=0]{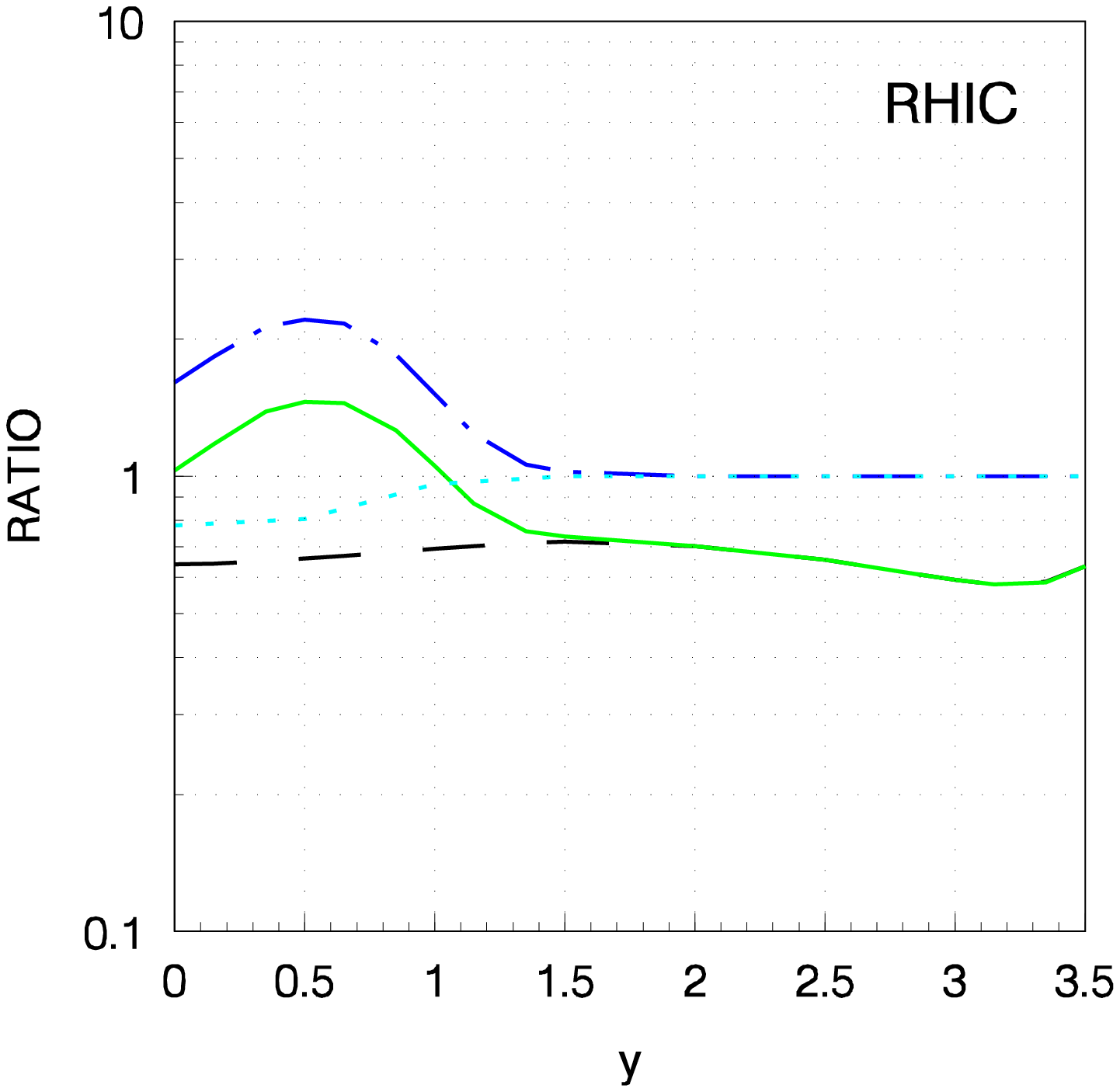}
  \end{center} 
\caption{The same as Fig. 13, except for rapidity distribution at $p_T=4$ 
GeV}  
\label{fig14} 
\end{figure} 

\newpage 
\begin{figure}[t] 
  \begin{center} 
     \includegraphics[width=0.7\textwidth,angle=0]{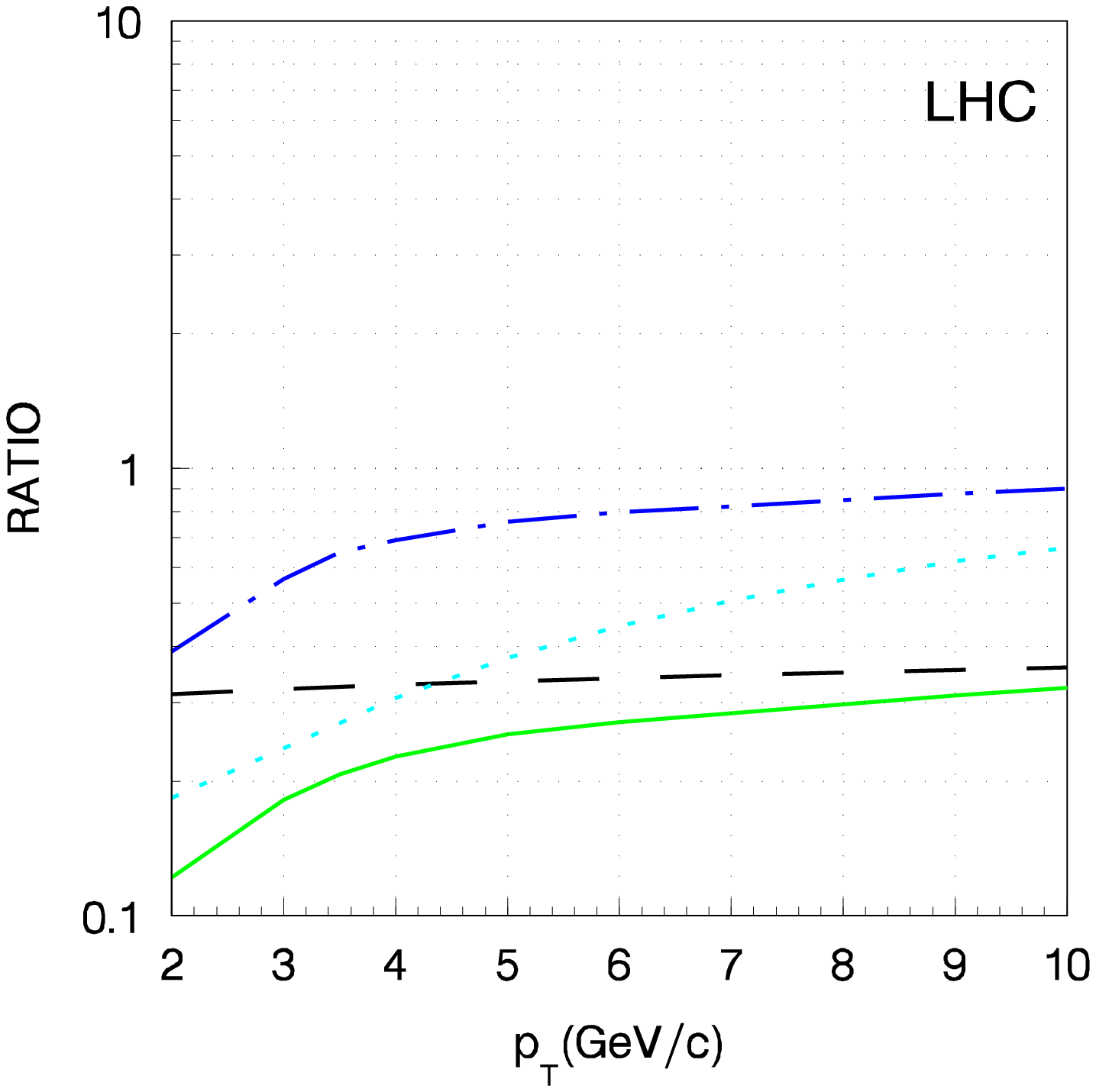}
  \end{center} 
\caption{Ratios versus transverse momentum at $y=0$ and LHC
energy. The solid, dashed, dot-dashed and dotted lines are
$R$, $R^{ini}$, $R^{plasma}$ and $S^{plasma}$, respectively.}  
\label{fig15} 
  \begin{center} 
     \includegraphics[width=0.7\textwidth,angle=0]{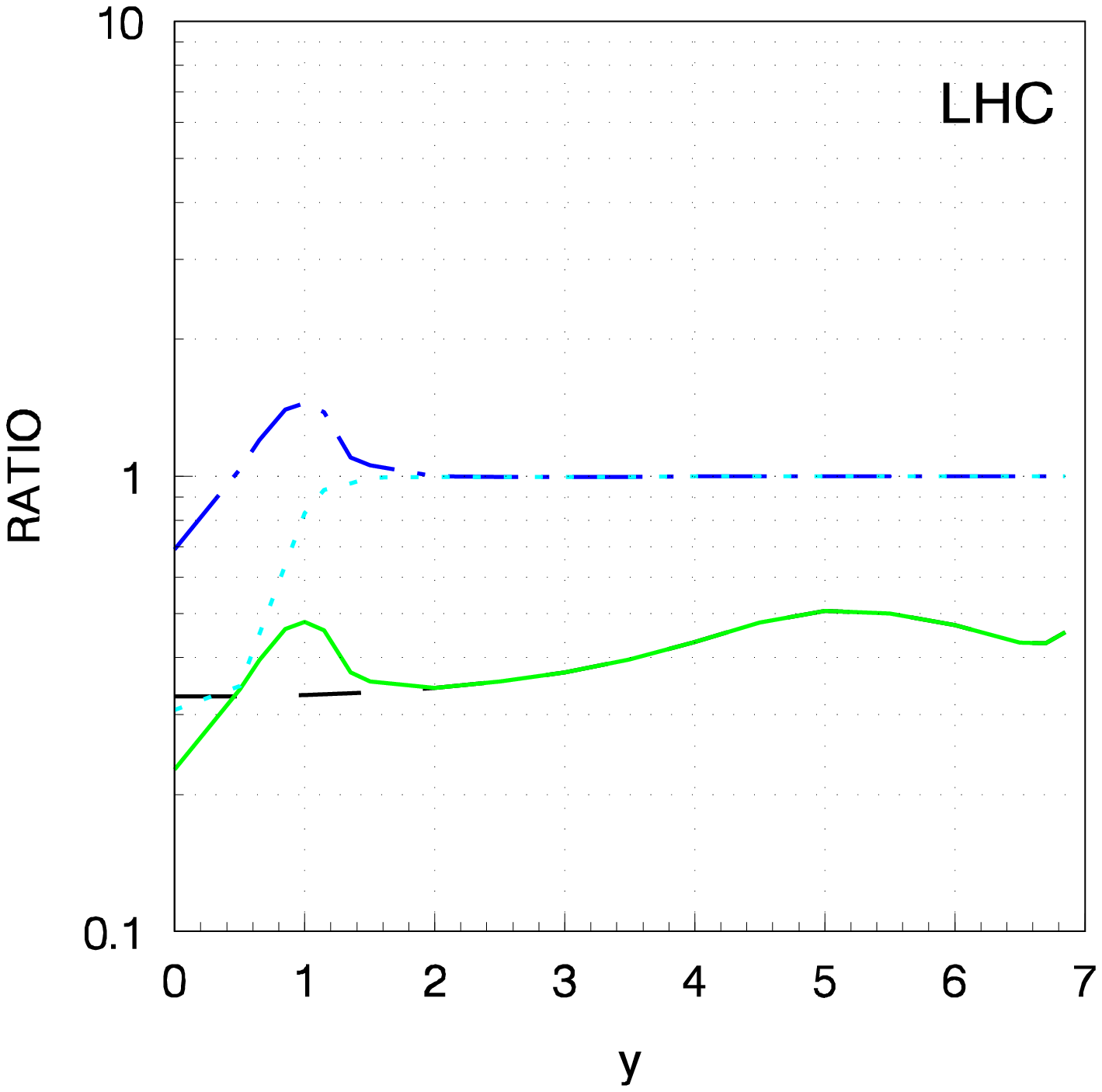}
  \end{center} 
\caption{The same as Fig. 15, except for rapidity distribution at 
$p_T=4$ GeV.}  
\label{fig16} 
\end{figure} 

\newpage 
\begin{figure}[t] 
  \begin{center} 
     \includegraphics[width=0.9\textwidth,angle=0]{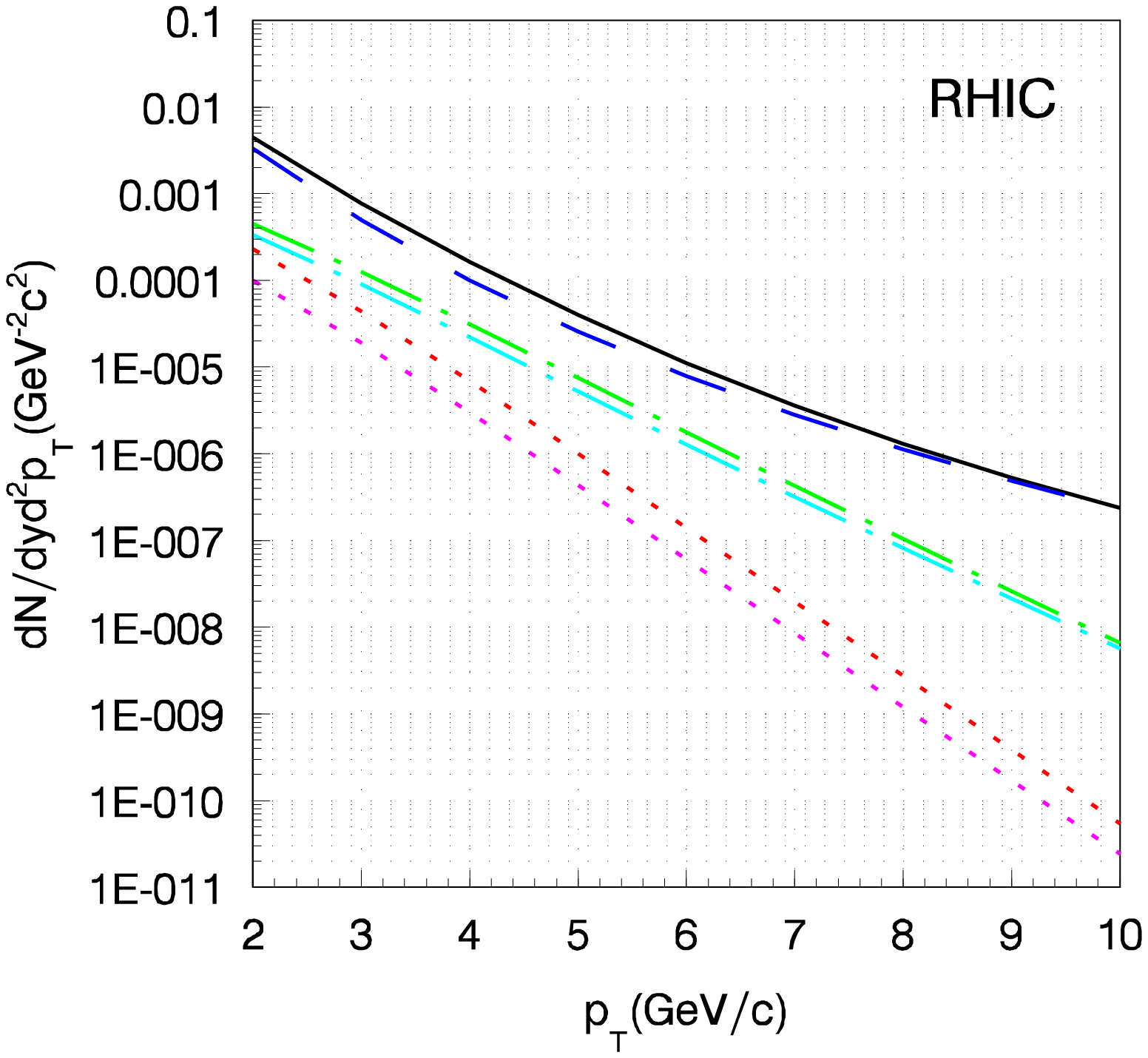}
  \end{center} 
\caption{$J/\psi$ number distributions versus transverse momentum at $y=0$ and
RHIC energy without suppression. The dashed curve corresponds to $c\bar c$ 
productions in the initial collision. The upper and lower
dot-dashed (dotted) curves correspond to $c\bar c$ produced through $2 \to 1$
and $2 \to 2$ reactions in the prethermal (thermal) stage, respectively.
The solid curve is the sum of all contributions.}  
\label{fig17} 
\end{figure} 

\newpage 
\begin{figure}[t] 
  \begin{center} 
     \includegraphics[width=0.9\textwidth,angle=0]{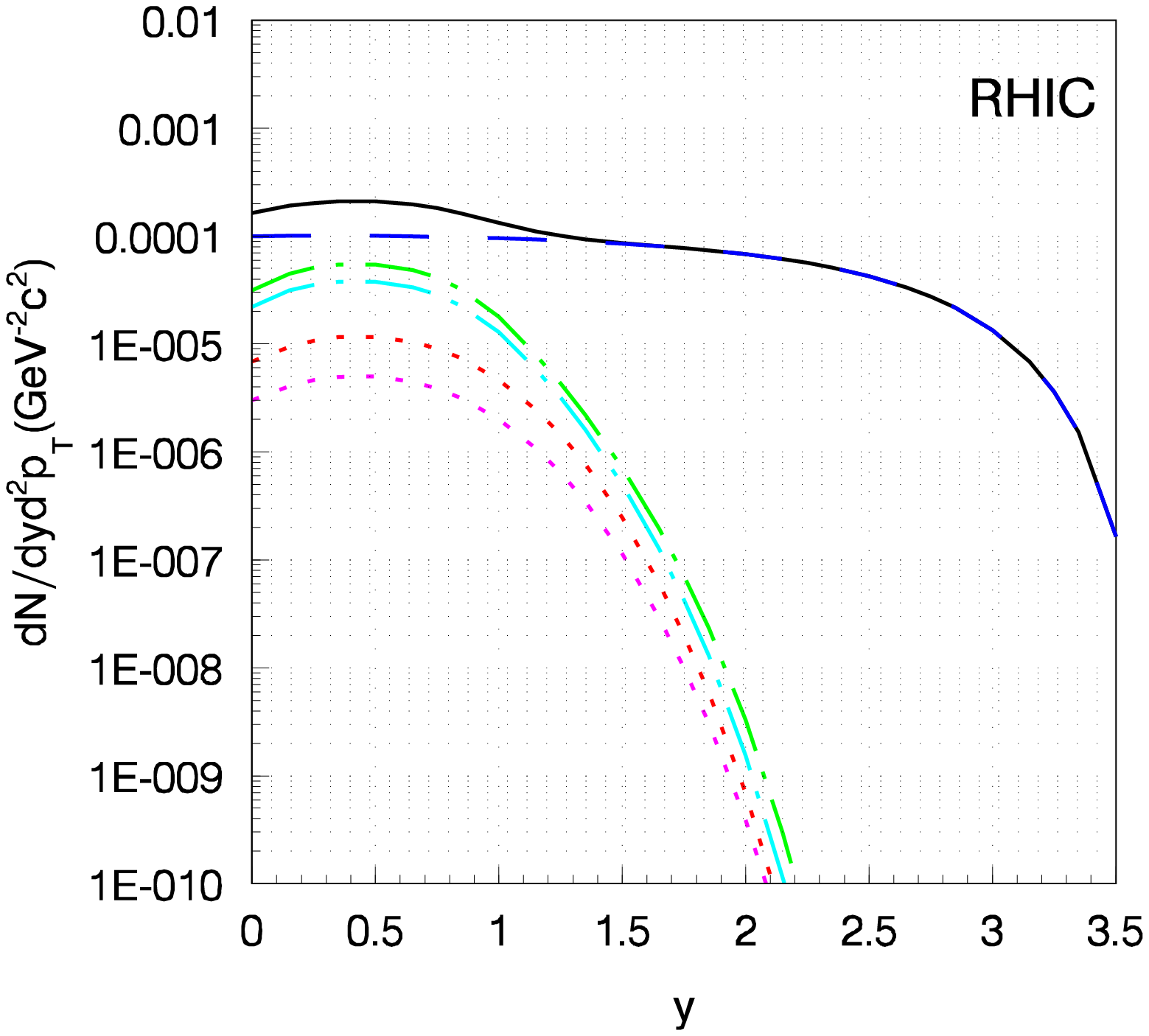}
  \end{center} 
\caption{The same as Fig. 17, except for rapidity distribution at
$p_T=4$ GeV.}  
\label{fig18}
\end{figure} 

\newpage 
\begin{figure}[t] 
  \begin{center} 
     \includegraphics[width=0.9\textwidth,angle=0]{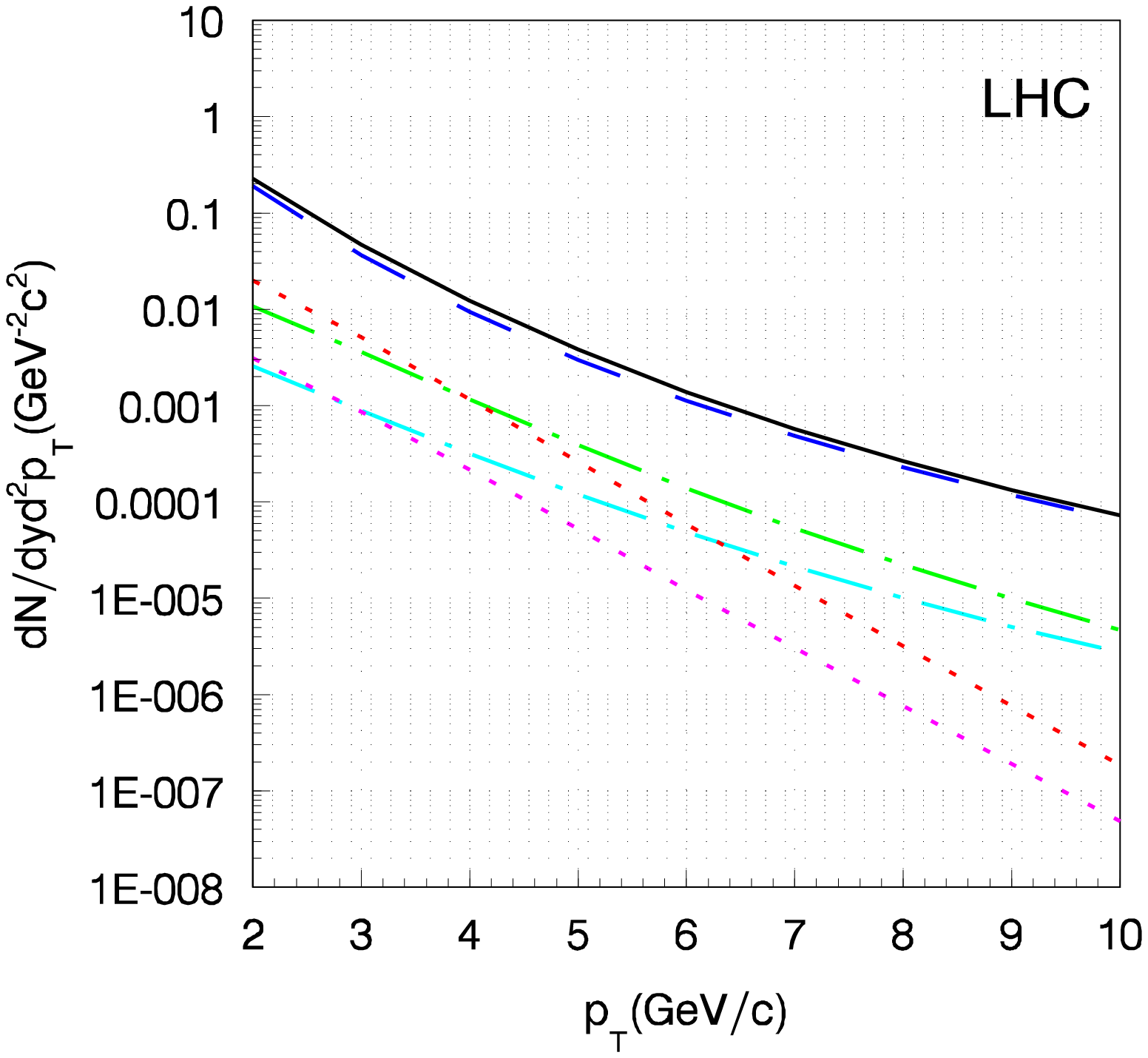}
  \end{center} 
\caption{$J/\psi$ number distributions versus transverse momentum at $y=0$ and
LHC energy without suppression. The dashed curve corresponds to $c\bar c$ 
productions in the initial collision. The upper and lower
dot-dashed (dotted) curves correspond to $c\bar c$ produced through $2 \to 1$
and $2 \to 2$ reactions in the prethermal (thermal) stage, respectively.
The solid curve is the sum of all contributions.}  
\label{fig19} 
\end{figure} 

\newpage 
\begin{figure}[t] 
  \begin{center} 
     \includegraphics[width=0.9\textwidth,angle=0]{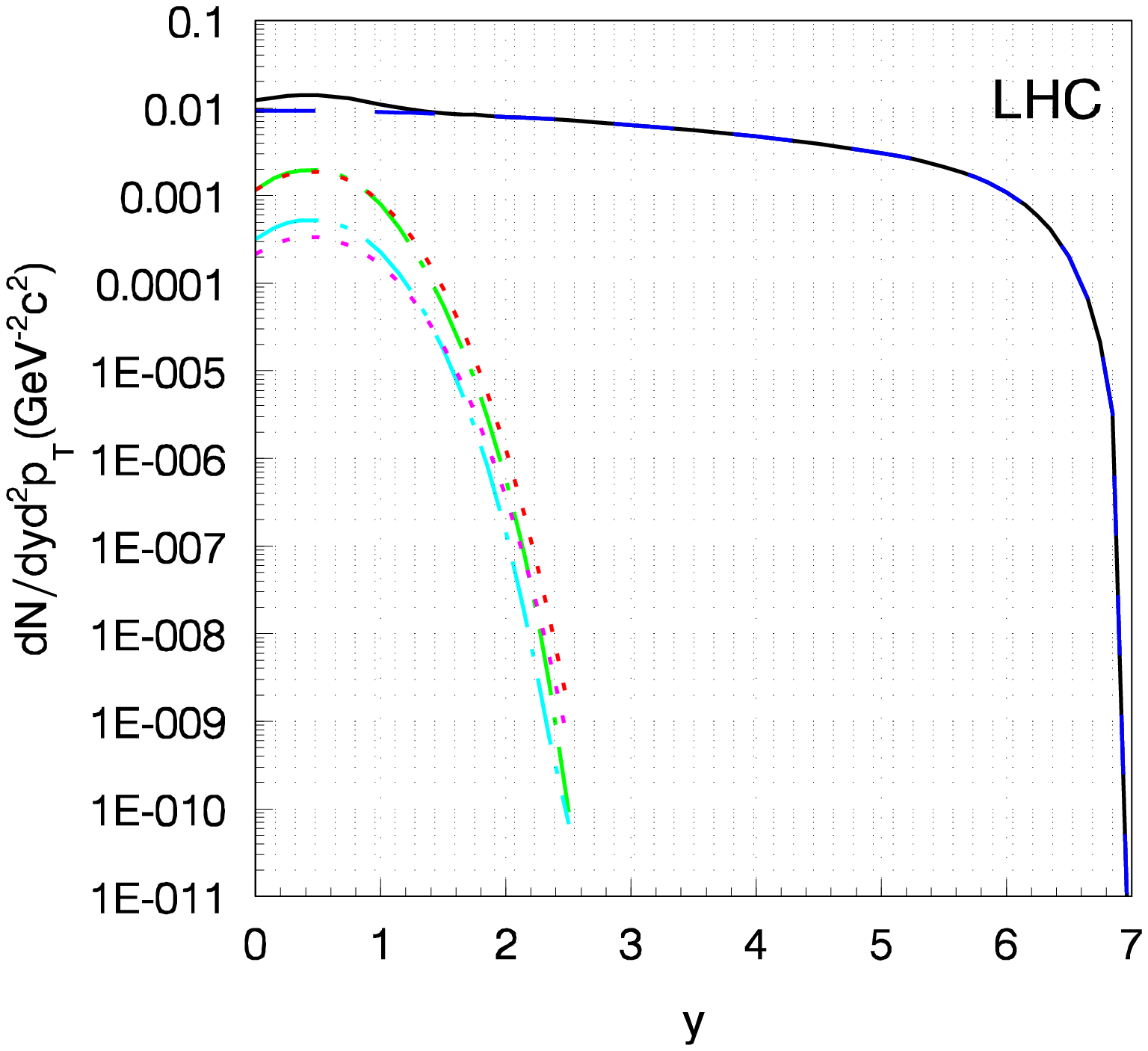}
  \end{center} 
\caption{ The same as Fig. 19, except for rapidity distribution at
$p_T=4$ GeV.}  
\label{fig20} 
\end{figure} 

\newpage 
\begin{figure}[t] 
  \begin{center} 
     \includegraphics[width=0.65\textwidth,angle=0]{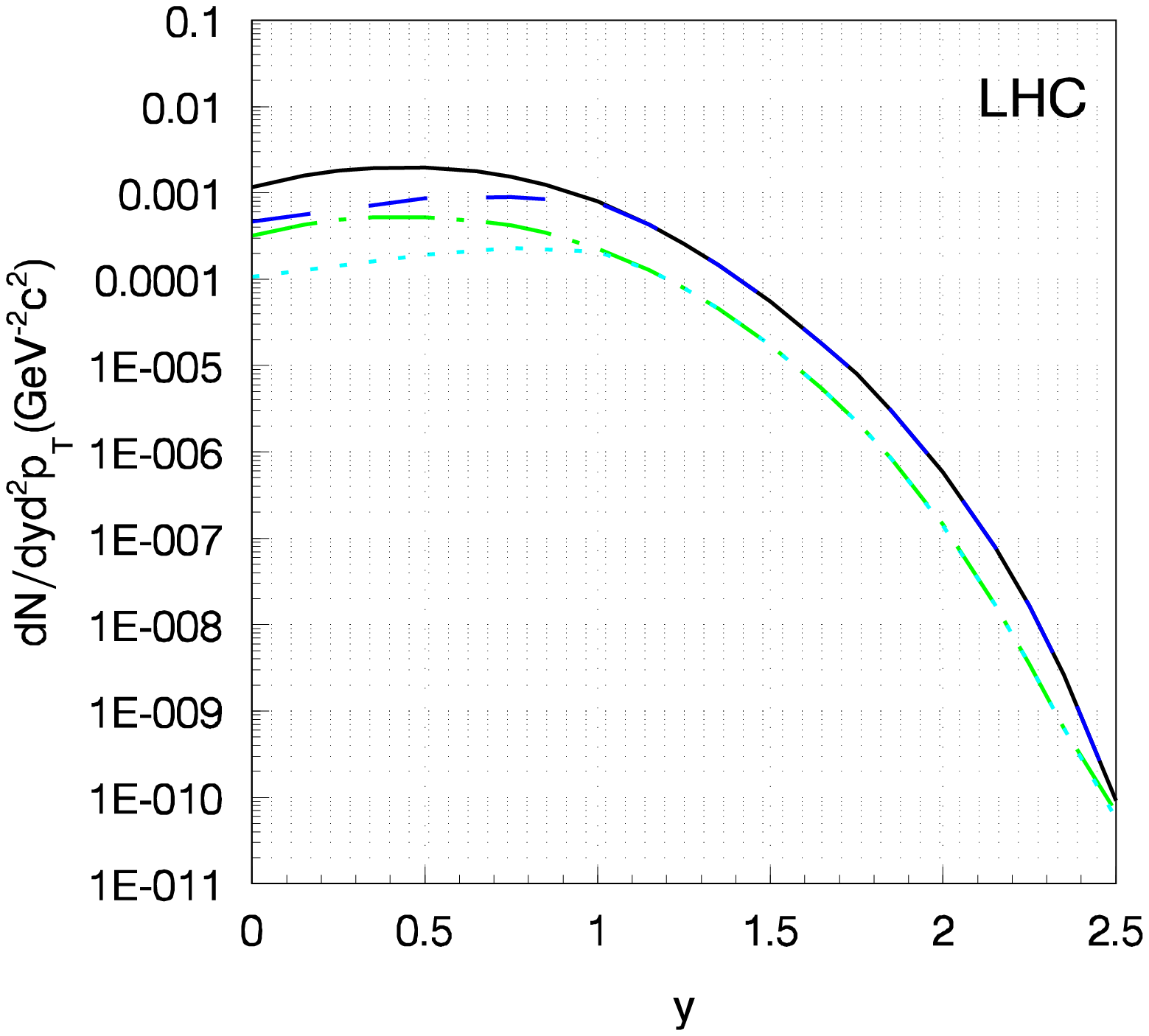}
  \end{center} 
\caption{$J/\psi$ number distributions  versus  rapidity at $p_T=4$ GeV in the
prethermal stage of LHC energy. The dashed and solid lines individually
correspond to $c\bar c$ produced through $2 \to 1$ reactions with and without 
suppression. The dotted and dot-dashed lines
through $2 \to 2$ reactions with and without suppression, respectively.}  
\label{fig21} 
  \begin{center} 
     \includegraphics[width=0.65\textwidth,angle=0]{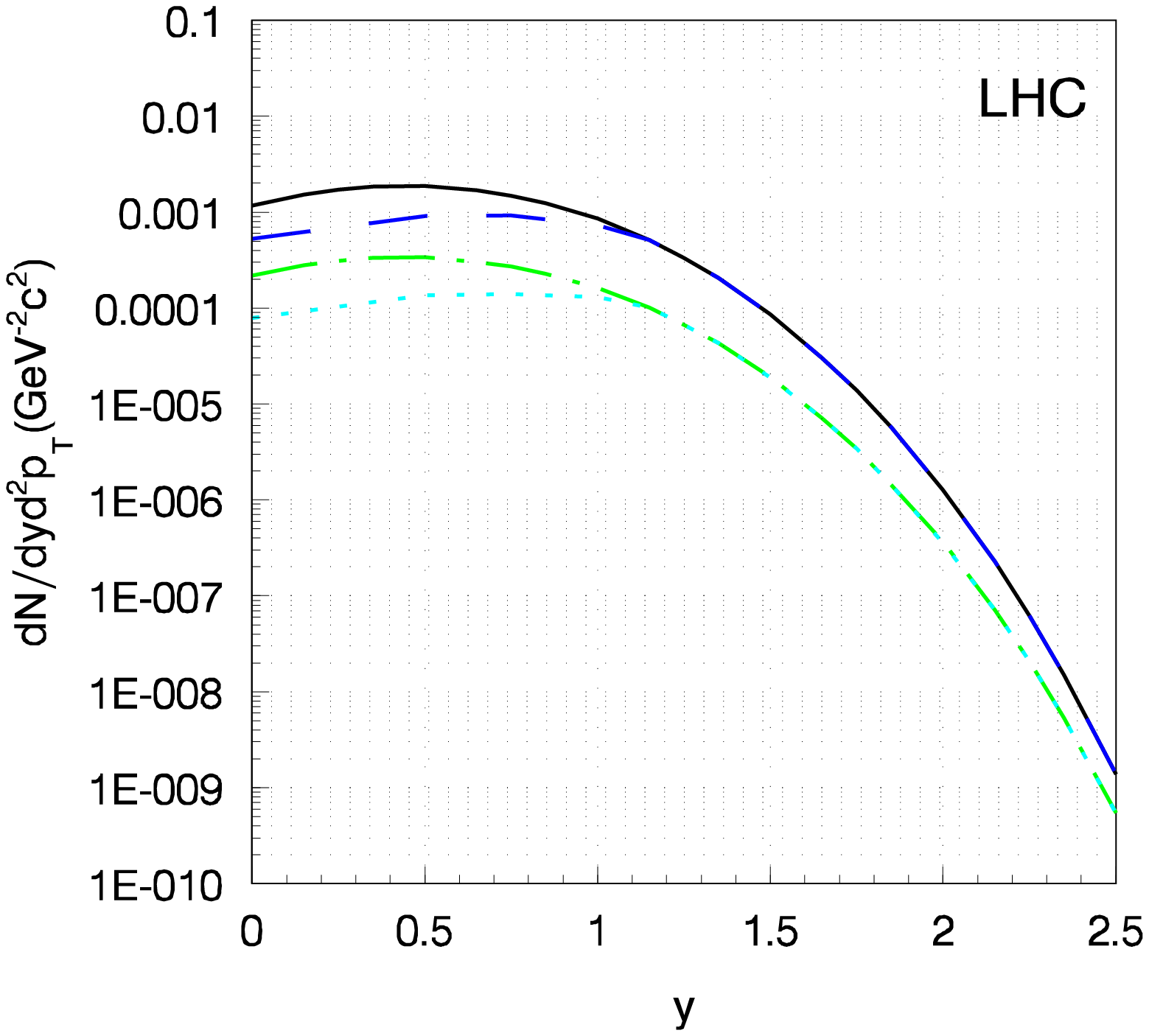}
  \end{center} 
\caption{The same as Fig. 21, except for the thermal stage.}  
\label{fig22} 
\end{figure} 

\newpage 
\begin{figure}[t] 
  \begin{center} 
     \includegraphics[width=0.7\textwidth,angle=0]{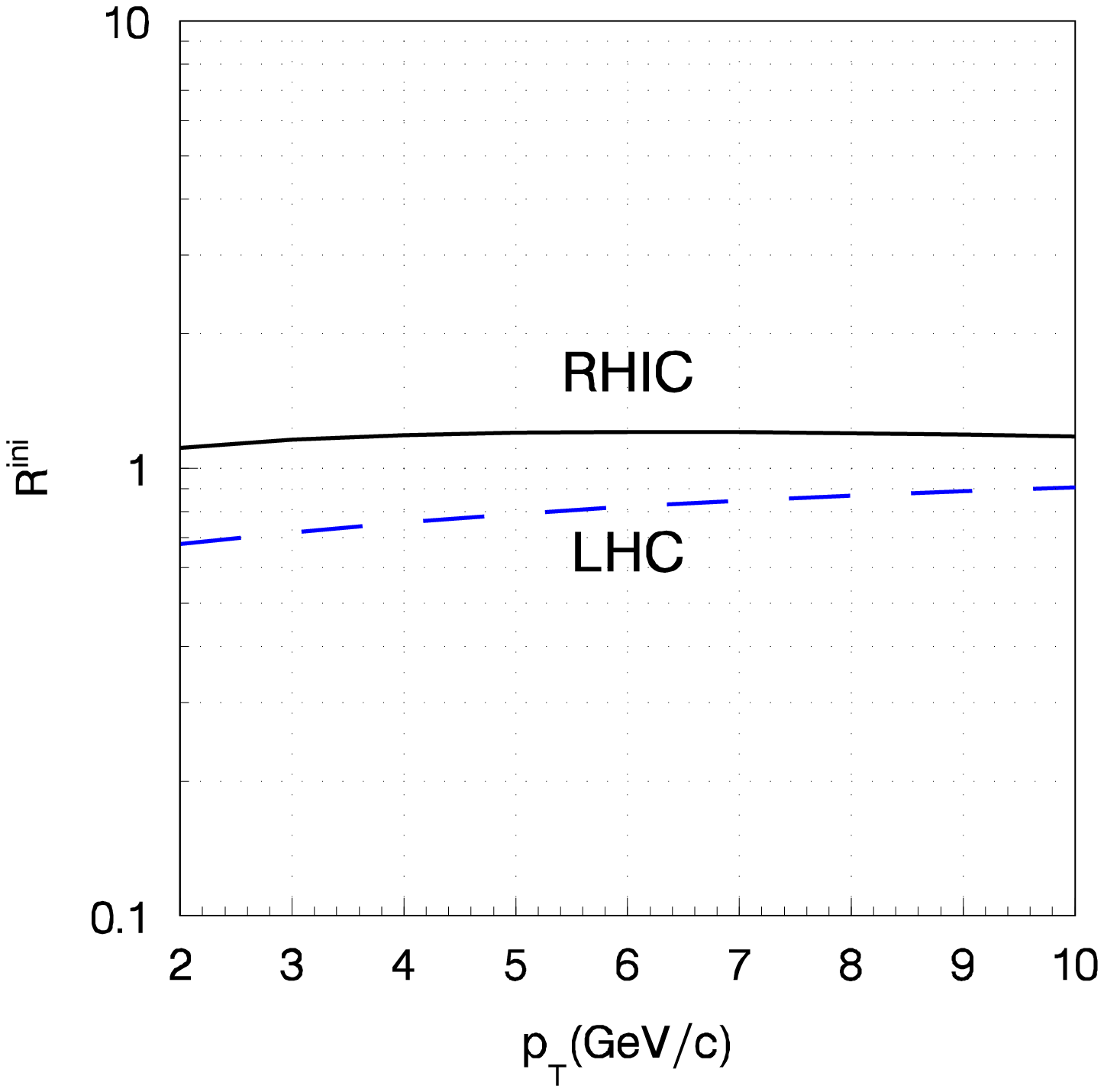}
  \end{center} 
\caption{Ratio $R^{ini}$ versus transverse momentum at $y=0$
is calculated with Eskola's parametrization.}  
\label{fig23} 
  \begin{center} 
     \includegraphics[width=0.7\textwidth,angle=0]{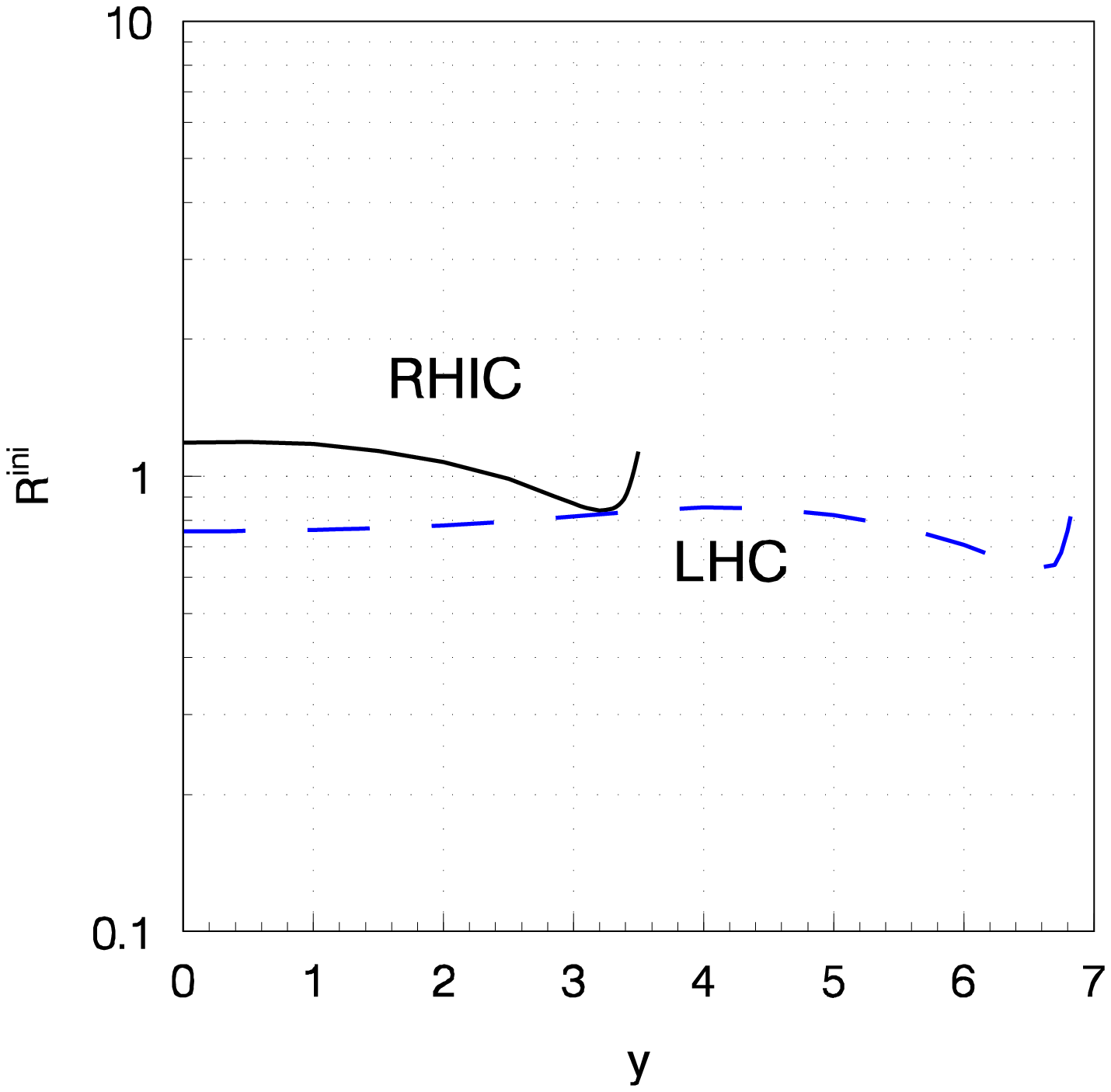}
  \end{center} 
\caption{The same as Fig. 23, except for rapidity distribution at 
$p_T=4$ GeV.}  
\label{fig24} 
\end{figure}

\end{document}